\documentclass[
 reprint,
superscriptaddress,
 amsmath,amssymb,
 aps,
pra,
floatfix,
]{revtex4-2}
\usepackage{graphicx}
\usepackage{dcolumn}
\usepackage{bm}
\usepackage[T1]{fontenc}
\usepackage{hyperref}
\DeclareMathOperator\erf{erf}

\begin{document}

\preprint{APS/123-QED}

\title{Engineering quantum droplet formation by cavity-induced long-range interactions}
\author{Leon Mixa}
\email{leon.mixa@uni-hamburg.de}
\affiliation{I. Institut für Theoretische Physik, Universität Hamburg, Notkestraße 9, 22607 Hamburg, Germany}
\affiliation{The Hamburg Center for Ultrafast Imaging, Luruper Chaussee 149, 22761 Hamburg, Germany}
\author{Milan Radonji\'c}
\affiliation{I. Institut für Theoretische Physik, Universität Hamburg, Notkestraße 9, 22607 Hamburg, Germany}
\affiliation{Institute of Physics Belgrade, University of Belgrade, Pregrevica 118,
11080 Belgrade, Serbia}
\author{Axel Pelster}
\affiliation{Physics Department and Research Center OPTIMAS, Rheinland-Pfälzische Technische Universität Kaiserslautern-Landau, Erwin-Schrödinger Straße 46,
67663 Kaiserslautern, Germany}
\author{Michael Thorwart}
\email{michael.thorwart@uni-hamburg.de}
\affiliation{I. Institut für Theoretische Physik, Universität Hamburg, Notkestraße 9, 22607 Hamburg, Germany}
\affiliation{The Hamburg Center for Ultrafast Imaging, Luruper Chaussee 149, 22761 Hamburg, Germany}
\date{\today}
\begin{abstract}
We investigate a dilute Bose gas with both a short-range contact and an effective long-range interaction between the atoms. The latter is induced by the strong coupling to a cavity light mode and is spatially characterized by a periodic signature and a tunable envelope rooted in the pumping of the cavity. We formulate a Bogoliubov theory based on a homogeneous mean-field description and quantum fluctuations around it.
The competition between the repulsive contact interaction and the long-range interaction allows the formation of self-bound quantum droplets. This generic approach is applied to two cavity setups, one without and one with a momentum-conserving effective long-range interaction between the atoms in the form of a driven dispersive cavity mode and a multimode cavity, respectively. For both cases we show analytically how the size and the central density of the cavity-induced quantum droplets depend on the contact interaction strength and on the shape of the spatial envelope of the long-range interaction.
\end{abstract}

\maketitle

\section{Introduction\label{introduction}}

Classical matter typically exists in one of three aggregate states: solid, liquid, or gas.
Their unique properties result from the interactions between their constituent particles.
At sufficiently low temperatures, quantum effects play a major role and enable more exotic  states of matter.
For example, below $2.17\;\!\rm K$, $^4$He exists as a superfluid \cite{Kapitza,Misener}, i.e., a quantum liquid that flows without friction and has no entropy.
In 1970 it was predicted that a new state, the supersolid, with both superfluid and solid properties should appear at very low temperatures \cite{Leggett}.
Although the experimental search for a supersolid has long focused on $^4$He, no conclusive supersolid properties have yet been discovered \cite{Chan}.

Helium, like a classical liquid, is capable of forming droplets, although it must be kept at low temperatures \cite{volovik2003universe}.
In the realm of ultracold dilute quantum gases, the existence of a new type of droplets was theoretically predicted by Petrov in Bose-Bose mixtures \cite{petrov2015quantum} and later experimentally verified \cite{cabrera2018quantum,semeghini2018self,skov2021observation}.
Unlike ordinary liquids, this quantum droplet state exhibits exceptionally low densities and is the result of an intricate interplay between weak mean-field attraction and repulsive quantum fluctuations.
The latter stabilize the gas, which would otherwise be unstable from a mean-field perspective.
A proper theoretical description of quantum droplets involves an extension of the mean-field Gross-Pitaevskii equation (GPe) by the famous Lee-Huang-Yang (LHY) correction \cite{LHY1957}.
As an alternative to homo- and heteronuclear atomic mixtures, a dipolar Bose-Einstein condensate (BEC) of magnetic atoms \cite{chomaz2022dipolar} can become unstable and break into multiple isolated self-bound quantum droplets when the dipolar interaction exceeds the contact interaction \cite{ferrier2016observation,pfau-nature}.
The corresponding dipolar extended GPe \cite{wachtler2016quantum,bisset2016ground} includes both an isotropic short-range contact interaction and an anisotropic long-range dipolar interaction, as well as the dipolar LHY correction \cite{schutzhold2006mean,lima2011quantum,lima2012beyond}.

An important advantage of dipolar BECs over their Bose-Bose mixture counterparts is the hosting of an unstable roton mode, which was predicted theoretically in 2003 \cite{roton-prediction1,roton-prediction2} and later verified experimentally \cite{chomaz2018observation,petter2019probing}.
It facilitates the spontaneous formation of a density pattern in a Bose superfluid and allows the formation of a supersolid \cite{tanzi2019observation,boettcher2019transient,chomaz2019long,donner2019dipolar,pelster2019supersolid}.
In the initial realizations of quantum droplets, their mutual distance was too large to establish global phase coherence among them, which is mandatory for a superfluid.
Only in 2019, three experiments with erbium and dysprosium atoms showed that phase-coherent quantum droplets and thus a supersolid exist in a very narrow parameter window as long as the number of atoms is sufficiently large \cite{tanzi2019observation,boettcher2019transient,chomaz2019long}.
In such a scenario, global phase coherence emerges precisely when the quantum droplets are connected by a background BEC.

Quantum droplet formation requires the presence and competition of two independent interactions in an ultracold atomic system.
When atoms are coupled to an optical cavity, a long-range interaction between them is effectively induced \cite{maschler2008ultracold}.
In addition, a roton mode is observed whose softness is controlled by the strength of the cavity-induced interaction \cite{mottl2012roton}.
When the roton mode becomes unstable, the system undergoes a quantum phase transition to a self-organized checkerboard density pattern accompanied by superradiance in the cavity \cite{baumann2010dicke}.

The long-range interaction in the cavity BEC system is qualitatively different from the dipolar one, which decays as $\propto r^{-3}$.
When talking about long-range interactions, the type with such an algebraic decay is commonly considered \cite{defenu2023long}.
The type of interaction is typically characterized by a stretched exponential decay of the form $\exp(-a r^{\alpha})$.
In view of recent proposals to engineer a variety of such interactions in cavity BEC systems \cite{bonifacio2024laser}, we present here a generic theory.
The effective interaction is realized by coupling of the BEC to a driven dispersive cavity mode, with the caveat that it is not momentum conserving.
In multimode cavities, a similar long-range interaction can be realized with a translation-invariant interaction potential.

We proceed as follows.
In Sec.\ \ref{classification} we discuss the criteria for a stable droplet and present a generic minimal model of the ground-state energy that can satisfy these conditions.
We identify three different classes of allowed parameters, which leads to the corresponding classification of quantum droplets.
In Sec.\ \ref{genericModel} we work out the homogeneous mean-field description and the quantum fluctuation correction around it for a generic system with both a contact and an effective long-range interaction between the atoms.
The latter is characterized by a spatially periodic signature that is modified by a tunable spatial envelope.
Based on these results, we discuss in Sec.\ \ref{factorizedResults} the case of a single-mode cavity coupled to a BEC, where the atoms are pumped by a transverse pump beam scattered into the cavity.
The transverse cavity mode profile naturally provides an envelope with the finite-range coupling to the cavity.
It thus leads to an effective long but finite-range interaction between the atoms, which we present in part in Ref.\ \cite{companionLetter}.
The general results for such a system are discussed in Sec.\ \ref{translationInvariantMultiCavity}. 
Finally, in Sec.\ \ref{conclusions} we summarize our findings on how the properties of cavity-induced quantum droplets can be tuned by both the contact interaction strength and the shape of the spatial envelope of the long-range interaction.

\section{Droplet classification\label{classification}}

The formation of a quantum droplet as a self-trapping quantum liquid that avoids self-evaporation relies on the fulfillment of three generic conditions originally introduced in the context of superfluid helium droplets \cite{volovik2003universe}:
\begin{equation}
\begin{aligned}
&\text{(C1) zero pressure}, \\
&\text{(C2) positive bulk compressibility}, \\
&\text{(C3) negative chemical potential}.
\end{aligned}\label{dropletConditions}
\end{equation}
The energy density of a minimal model that satisfies these conditions contains two terms.
The first term is usually due to the mean-field energy and depends quadratically on the density $n$, while the second term also depends algebraically on the density with a power exponent that differs from the former by the parameter $\gamma$. 
For example, a three-dimensional Bose-Bose mixture and a dipolar Bose gas allow for a distinct class of quantum droplets where the mean-field term is attractive and repulsive quantum fluctuations are characterized by the exponent $5/2$, such that $\gamma = 1/2$.
A second class of quantum droplets occurs in a one-dimensional Bose-Bose mixture where, conversely, the mean-field contribution is positive and the quantum corrections yield a negative one with the exponent parameter $\gamma = -1/2$ \cite{petrov2016ultradilute}.

In this work, we theoretically introduce a third class of quantum droplets.
They should arise in systems, such as those realizable in cavity BEC experiments, that exhibit specific long-range interactions.
In such a scenario, the quantum fluctuations of the cavity give rise to the roton mode.
Unlike for the other droplet types, the corresponding energy density cannot be expressed as a function of atomic density alone.
Consequently, when applying the above droplet conditions \eqref{dropletConditions}, we cannot rely on the study of the system energy density.
However, as we will show, it is possible to generalize these conditions to our finite-size system by using the corresponding effective 
ground-state energy as the basis for a generic minimal model.
To this end, we assume that the number $N$ of atoms is fixed and consider the effective ground-state energy of a self-bound quantum liquid in the form
\begin{align}
E_0(N,V) = \frac{\alpha(N)}{V} + \frac{\beta(N)}{V^{1+\gamma}} + E_\mu(N)  \, .
\label{minimalModel}
\end{align}
Importantly, here we permit a system-size-independent energy contribution $E_\mu(N)$, which does not affect the equilibrium size, but only influences the chemical potential.
For a self-bound droplet, the energy $E_0$ must have a minimum at a certain system volume $V_0(N) > 0$ (C1), which implies
\begin{align}
\left( \frac{\partial E_0}{\partial V} \right)_N \bigg|_{V=V_0} \overset{!}{=} 0 \; \Rightarrow \; V_0(N)^{\gamma} = - \frac{(1+\gamma)\beta(N)}{\alpha(N)} \, .
\label{condition1}
\end{align}
In order that this local extremum is actually a minimum, we must demand (C2)
\begin{align}
\left( \frac{\partial^2 E_0}{\partial V^2} \right)_N \bigg|_{V=V_0} \overset{!}{>} 0 \hspace*{0.5cm} \Rightarrow \hspace*{0.5cm} \gamma \alpha < 0 \, .
\label{condition2}
\end{align}
The combination of (\ref{condition1}) and (\ref{condition2}) finally leads to three possible classes of parameters $\alpha$, $\beta$ and $\gamma$ that can realize a quantum droplet:
\begin{equation}
\begin{aligned}
&\text{(D1) } \alpha<0 \, , \, \beta>0 \, , \, \gamma >0 \, , \\
&\text{(D2) } \alpha>0 \, , \,  \beta<0 \, , \, 0>\gamma>-1 \, , \\
&\text{(D3) } \alpha>0 \, , \, \beta>0 \, , \, \gamma<-1 \, .
\end{aligned}
\label{dropletClasses}
\end{equation}
The analog of the third droplet condition (C3), by virtue of (C1), becomes
\begin{eqnarray}
\hspace*{-4mm}\bigg(\frac{\partial E_0}{\partial N}\bigg)_{V=V_0} = \frac{1}{V_0} \frac{d\alpha}{dN} + \frac{1}{V_0^{1+\gamma}} \frac{d\beta}{dN} + \frac{dE_\mu(N)}{dN} \overset{!}{<} 0 \, ,
\end{eqnarray}
which yields
\begin{eqnarray}
 \frac{d\alpha}{dN} - \frac{\alpha}{(1+\gamma)\beta} \frac{d\beta}{dN}+ \frac{dE_\mu(N)}{dN} < 0 \, . \hspace{5mm}
\label{condition3}
\end{eqnarray}
This must be checked for the specific $N$ dependence of $\alpha(N)$, $\beta(N)$, and the energy shift $E_\mu(N)$.

\begin{figure}
\centering\includegraphics[scale=.3]{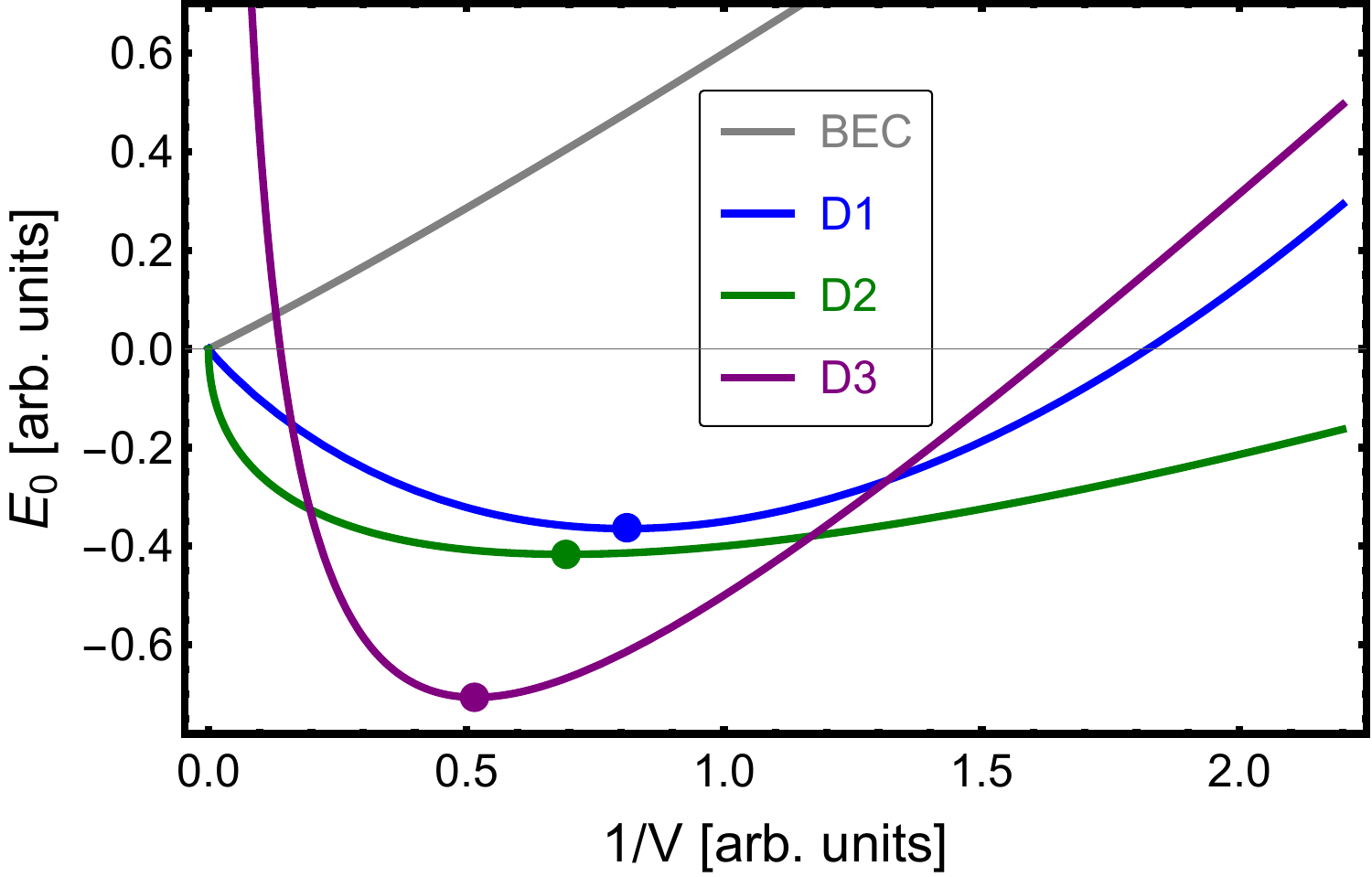}
\vspace{-1mm}
\caption{Sketch of the effective ground-state energy $E_0$ of the model Eq.\ (\ref{minimalModel}) against the inverse system size $1/V$ for a weakly interacting BEC and the three droplet classes of Eq.\ (\ref{dropletClasses}). The energy minima at the equilibrium volume $V_0$, obtained from Eq.\ (\ref{condition1}), are indicated by dots. The sketch parameters are: $\alpha = 0.5$, $\beta = 0.01$, $\gamma = 1/2$, (D1) $\alpha = -1.35$, $\beta = 1$, $\gamma = 1/2$, (D2) $\alpha = 0.6$, $\beta = -1$, $\gamma = -1/2$, and (D3) $\alpha = 1$, $\beta = 1/2$, $\gamma = -5/3$. The curve for (D3) includes a system size independent shift $E_\mu$.\vspace{-2mm}\label{minimalModelPressure}}
\end{figure}

In Fig.\ \ref{minimalModelPressure} we sketch the two droplet conditions (C1) and (C2) of Eq.\ \eqref{dropletConditions} for a BEC and the three possible droplet classes (D1)--(D3).
A simple BEC with only repulsive weak contact interaction has an effective energy $E_0=gN^2/2V$, which grows as $\propto 1/V$ with the inverse system size.
In the infinitely dilute limit, its ground-state energy is zero.
The established dipolar droplets and Bose-Bose mixture droplets in three dimensions belong to the class (D1) defined in Eq.\ (\ref{dropletClasses}).
In the dilute limit $1/V \to 0$ the unstable attractive mean field $\propto - 1/V$ dominates.
Figure \ref{minimalModelPressure} shows that the interplay with its positive repulsive quantum correction $\propto 1/V^{3/2}$, which dominates in the dense limit, leads to a minimum of the effective energy.
Thus, a stable droplet is formed between these limits for finite system size and density.
One-dimensional Bose-Bose mixtures, on the other hand, belong to the class (D2).
They form droplets through the interplay of the attractive quantum fluctuation contribution $\propto -1/V^{1/2}$ and the repulsive mean-field term $\propto 1/V$, so that the energetic minimum is seen in the green plot of Fig.\ \ref{minimalModelPressure}.

We will show in this work that the long-range interactions engineered in cavity BEC systems realize the third droplet class (D3).
In the limit $1/V \to 0$ the effective energy $E_0$ follows a power law $\propto 1/V^{1+\gamma}$ due to the quantum correction term. In Fig.\ \ref{minimalModelPressure} we choose $\gamma=-5/3$ based on our results in Sec.\ \ref{factorizedResults}.
In the physical systems discussed below, the divergence of $E_0$ in the infinitely dilute limit is an artifact of the validity of the approximations we use in the derivation.
In the limit $1/V \to \infty$, the repulsive mean field determines the effective ground-state energy with the growth $\propto 1/V$.
For this effective energy model of type (D3) to satisfy the droplet condition (C3), the presence of a sufficiently negative size-independent energy $E_\mu(N)$ is required, which occurs naturally in the systems we study in this work.

\section{Generic model \label{genericModel}}

We study an effective generic model of a dilute gas of bosonic atoms (of individual mass $M$) in $d$ dimensions in a cavity. The atoms experience two different types of density-density interactions: a contact interaction of strength $g$ and an effective long-range interaction mediated by the cavity.
The effective Hamiltonian is (with $\hbar = 1$)
\begin{align}
&\hat{H}_{\text{eff}} = \int_V d^d\bm{r} \, \hat{\psi}^{\dag}(\bm{r})  \left[ -\frac{\mbox{$\bm{\nabla}^2$}}{2M}  + \frac{g}{2} \,\hat{\psi}^{\dag}(\bm{r}) \hat{\psi}(\bm{r}) \right]\hat{\psi}(\bm{r}) \nonumber \\
&+{}\frac{1}{2} \int_V d^d\bm{r}\int_V d^d \bm{r}' \hat{\psi}^\dag(\bm{r}) \hat{\psi}(\bm{r}) V_{{\rm C}}(\bm{r},\bm{r}') \hat{\psi}^{\dag}(\bm{r}') \hat{\psi}(\bm{r}') \, ,
\label{effectiveInteractingBEC}
\end{align}
where $\hat{\psi}$, $\hat{\psi}^\dagger$ are bosonic field operators and the long-range interaction potential
\begin{align}
V_{{\rm C}}(\bm{r},\bm{r}') = \mathcal{I} v(\bm{r},\bm{r}') f_{\bm{\xi}}(\bm{r},\bm{r}') \, ,
\label{longRangePotential}
\end{align}
has three real-valued constituents. The interaction strength is denoted by $\mathcal{I}$ and $v(\bm{r},\bm{r}')$ stands for a dimensionless periodic function, obeying $|v(\bm{r},\bm{r}')| \leq 1$ as well as the symmetry $v(\bm{r},\bm{r}')=v(\bm{r}',\bm{r})$.
Furthermore, $f_{\bm{\xi}}(\bm{r},\bm{r}')$ represents a dimensionless envelope with widths $\bm{\xi}$, which is characterized by the properties $f_{\bm{\xi}}(\bm{0},\bm{0}) = 1$, $ |f_{\bm{\xi}}(\bm{r},\bm{r}')| \leq 1$, $\lim_{|\bm{\xi}| \to \infty} f_{\bm{\xi}}(\bm{r},\bm{r}') = 1$, and the symmetry $f_{\bm{\xi}}(\bm{r},\bm{r}') = f_{\bm{\xi}}(\bm{r}',\bm{r})$.
Below in Secs.\ \ref{factorizedResults} and \ref{translationInvariantMultiCavity} we consider two different experimental setups, described effectively by the Hamiltonian (\ref{effectiveInteractingBEC}) and by a long-range interaction potential of the form (\ref{longRangePotential}).
This justifies the following analysis of useful properties of the interaction potential (\ref{longRangePotential}) and the subsequent development of a Bogoliubov theory for the effective Hamiltonian (\ref{effectiveInteractingBEC}).
The resulting ground-state energy will then be used to study the formation of cavity-induced droplets.

\subsection{Fourier Transformation}

The atoms occupy a space of finite extent $L_{\nu}$ in each direction $\nu = 1,\dots,d$, so the volume of the system is $V = \prod_{\nu = 1}^d L_{\nu}$.
Since we are not concerned here with edge effects, we assume periodic boundary conditions in each of the directions.
Therefore, the field operators can be expanded in their respective Fourier series
\begin{align}
\hat{\psi}(\bm{r}) = \frac{1}{\sqrt{V}} \sum_{\bm{p}} e^{i\bm{p}\bm{r}}\:\! \hat{\psi}_{\bm{p}} \, ,
\end{align}
with the Fourier amplitudes given by
\begin{align}
\hat{\psi}_{\bm{p}} = \frac{1}{\sqrt V} \int_{V} d^d \bm{r}\:\! e^{-i\bm{p}\bm{r}} \:\! \hat{\psi}(\bm{r}) \, ,
\end{align}
where the momentum components take the discrete values $p_{\nu}=2\pi m_{\nu}/L_{\nu}$ with integer $m_{\nu}$.
We choose the origin of the coordinate system such that the $\bm{\lambda}$-periodic potential $v(\bm{r},\bm{r}')$ is an even function, i.e., $v(\bm{r},\bm{r}') = v(-\bm{r},-\bm{r}')$.
The periods $\lambda_{\nu}$ are integer fractions of the system extension $L_{\nu}$ in each direction, i.e., we have $\lambda_{\nu} = L_{\nu}/l_{\nu}$ for $l_{\nu} \in \mathbb{Z} \setminus \lbrace 0 \rbrace$.
Consequently, the periodic potential $v(\bm{r},\bm{r}')$ can be represented by its Fourier series 
\begin{align}
v(\bm{r},\bm{r}') = \sum_{\bm{k},\bm{k}'}  e^{i\bm{k}\bm{r}+i\bm{k}'\bm{r}'}\:\!\tilde{v}_{\bm{k},\bm{k}'} \, ,
\label{Fourier}
\end{align}
with the Fourier amplitudes given by
\begin{align}
\tilde{v}_{\bm{k},\bm{k}'} = \int_{V} \frac{d^d \bm{r}}{V} \int_{V} \frac{d^d\bm{r}'}{V}\:\! e^{-i\bm{k}\bm{r} - i\bm{k}'\bm{r}'} v(\bm{r},\bm{r}') \, ,
\end{align}
with $k_\nu = 2\pi j_\nu/\lambda_\nu$, $k_\nu' = 2\pi j_\nu'/\lambda_\nu$ and $j_\nu,j_\nu' \in \mathbb{Z}$.
Note that the set of wave vectors $\bm{k}$, denoted as $\mathcal{K}_{\rm C}$, is a proper sublattice of the set of momenta $\bm{p}$, since the respective periods are commensurable.
We will exclude $\bm{k}=\bm{0}$ from $\mathcal{K}_{\rm C}$, which will turn out to be necessary to have a purely beyond mean-field effect of the long-range interaction.
Physically, this means that the periodic potential $v(\bm{r},\bm{r}')$ does not have a constant baseline, neither in $\bm{r}$ nor in $\bm{r}'$, i.e., $\tilde{v}_{\bm{k},\bm{0}}=\tilde{v}_{\bm{0},\bm{k}'}=0$.
Below, the introduction of the expansion \eqref{Fourier} into various spatial integrals over the volume $V$ will lead to the selection of the Fourier amplitudes of the rest of the integrand, corresponding to the set $\mathcal{K}_{\rm C}$, which will be the only ones that result in a non-zero contribution.

The above symmetry properties of $v(\bm{r},\bm{r}')$ translate to the Fourier amplitudes via
\begin{align}
\tilde{v}_{\bm{k},\bm{k}'} = \tilde{v}_{\bm{k}',\bm{k}} = \tilde{v}_{-\bm{k},-\bm{k}'} \, .
\label{v-FT-properties}
\end{align}
Note that we do not impose translational invariance on the long-range potential (\ref{longRangePotential}), since we are also interested in potentials realized as effective interactions mediated by dissipative bosonic modes, which do not necessarily conserve momentum.

To continue, we need the Fourier expansion of the envelope function
\begin{align}
f_{\bm{\xi}}(\bm{r},\bm{r}') = \sum_{\bm{p},\bm{p}'} e^{i\bm{p}\bm{r} + i\bm{p}'\bm{r}'} \tilde{f}_{\bm{\xi}}(\bm{p},\bm{p}') \, ,
\label{FourierEnv}
\end{align}
where the Fourier amplitudes read
\begin{align}
\tilde{f}_{\bm{\xi}}(\bm{p},\bm{p}') = \int_V \frac{d^d\bm{r}}{V} \int_V \frac{d^d\bm{r}'}{V}\:\! e^{-i\bm{p}\bm{r} - i\bm{p}'\bm{r}'} f_{\bm{\xi}}(\bm{r},\bm{r}') \, .
\label{envelopeFourierSeries}
\end{align}
First, we note that $\tilde{f}_{\bm{\xi}}(\bm{0},\bm{0})$ is the spatial average of the dimensionless envelope $f_{\bm{\xi}}(\bm{r},\bm{r}')$.
In the following we assume that $f_{\bm{\xi}}(\bm{r}, \bm{r}')$ varies only weakly on the scale of the system size, so that it can be well approximated by its spatial average $\tilde{f}_{\bm{\xi}}(\bm{0},\bm{0})$ next to the complex exponentials in Eq.\ (\ref{envelopeFourierSeries}) and then taken out of the integrals.
In this way, we arrive at
\begin{align}
\tilde{f}_{\bm{\xi}}({\bm p},{\bm p}') \approx \delta_{{\bm p}\bm{0}}^{(d)} \delta_{{\bm p}'\bm{0}}^{(d)} \:\!\tilde{f}_{\bm{\xi}}(\bm{0},\bm{0}) \, .
\label{approxKroneckerDelta}
\end{align}
Thus, $\tilde{f}_{\bm{\xi}}(\bm{0},\bm{0})$ is the only envelope property relevant to the discussion in the following.
    
\subsection{Mean-Field}

Motivated by the preceding section, we analyze the model Hamiltonian (\ref{effectiveInteractingBEC}) within the framework of Bogoliubov theory. For this purpose, we employ the ansatz 
\begin{align}
\label{ansatz}
\hat{\psi}(\bm{r}) =\sqrt{n}  + \hat{\phi}(\bm{r})\, ,
\end{align}
where the first term denotes the homogeneous mean-field with the particle density $n = N/V$ and 
the effect of the quantum fluctuations around this mean-field are described by the fluctuation operator $\hat{\phi}(\bm{r})$.
We start by neglecting the quantum fluctuations, so that the first line in the effective Hamiltonian Eq.\ (\ref{effectiveInteractingBEC}) gives straightforwardly the atomic mean-field energy
\begin{align}
E_{\rm mf,A} = \frac{gn^2}{2}V \, .
\end{align}
The double integral in the second line of Eq.\ (\ref{effectiveInteractingBEC}) yields the cavity-induced contribution
\begin{align}
E_{\rm mf,C}=\frac{\mathcal{I}n^2}{2}  \int_V d^d\bm{r} \int_V d^d\bm{r}' v(\bm{r},\bm{r}') f_{\bm{\xi}}(\bm{r},\bm{r}')
\, .
\end{align}
Taking into account the Fourier series in Eq.\ (\ref{Fourier}) of the periodic function  $v(\bm{r},\bm{r}')$ and  (\ref{approxKroneckerDelta}) yields
\begin{align}
E_{\rm mf,C} &= \frac{\mathcal{I}N^2}{2} \tilde{v}_{\bm{0},\bm{0}} \tilde{f}_{\bm{\xi}}(\bm{0},\bf{0})  \, .
\label{longRangeMeanField}
\end{align}
Thus, the long-range interaction contributes to the homogeneous mean-field only if $\tilde{v}_{\bm{0},\bm{0}} \neq 0$ and $\tilde{f}_{\bm{\xi}}(\bm{0},\bm{0}) \neq 0$. 
In the following, we are interested in going beyond mean-field effects in a homogeneous system.
Therefore we will continue with the examination of long-range interactions without the spatially constant background and assume $\tilde{v}_{\bm{0},\bm{0}} = 0$.
In this case we have $E_{\rm mf,C} = 0$, so the mean-field of the system is unaffected by the long-range interaction $V_{\rm C}$, and the mean-field chemical potential is given by $\mu_{\rm mf} = gn$.

\subsection{First-Order Quantum Fluctuations}

To determine the effect of quantum fluctuations, we Fourier-expand the fluctuation operator
\begin{align}
\label{Fourier2}
{\hat{\phi}(\bm{r}) = \frac{1}{\sqrt{V}}}\sum_{\bm{p}}{}^{'} e^{i\bm{p}\bm{r}}\:\! \hat{\phi}_{\bm{p}} \,,  
\end{align}
where the primed sum denotes omitting the $\bm{p}=\bm{0}$ term. Inserting the ansatz of Eq.\ (\ref{ansatz}) into the model Hamiltonian Eq.\ (\ref{effectiveInteractingBEC}) yields in first order
\begin{eqnarray}
\hat{H}_1 &=& \mathcal{I} n^{3/2} \int_V d^d\bm{r} \big[\hat{\phi}(\bm{r}) + \hat{\phi}^{\dag}(\bm{r})\big] \nonumber\\
&&\times\int_V d^d\bm{r}' v(\bm{r},\bm{r}') f_{\bm{\xi}}(\bm{r},\bm{r}')\, .
\end{eqnarray}
Due to the Fourier expansions \eqref{Fourier}, \eqref{FourierEnv} and \eqref{Fourier2} as well as the approximation (\ref{approxKroneckerDelta}), this leads to the result
\begin{align}
\hat{H}_1&= \mathcal{I}N^{3/2} \tilde{f}_{\bm{\xi}}(\bm{0},\bm{0}) \sum_{\bm{k} \in \mathcal{K}_{\rm C}} \big( \hat{\phi}_{-\bm{k}} + \hat{\phi}_{\bm{k}}^{\dagger} \big) \tilde{v}_{\bm{k},\bm{0}} \, ,
\label{H1}
\end{align}
which vanishes due to the assumed absence of the constant baseline of $v(\bm{r},\bm{r}')$.

\subsection{Second-Order Quantum Fluctuations}

Let us next turn to the second-order effect of the quantum fluctuations.
The corresponding part of the Hamiltonian Eq.\ (\ref{effectiveInteractingBEC}) reads
\begin{eqnarray}
\hat{H}_2 &=& \frac{1}{2} \sum_{\bm{p}}{}^{'} \bigg[ \frac{\bm{p}^2}{2M} \big( \hat{\phi}_{\bm{p}}^{\dagger} \hat{\phi}_{\bm{p}} + \hat{\phi}_{-\bm{p}} \hat{\phi}_{-\bm{p}}^{\dagger} \big) - \frac{\bm{p}^2}{2M} \nonumber \\
&&{}+ gn \big( \hat{\phi}_{-\bm{p}} + \hat{\phi}_{\bm{p}}^{\dagger} \big) \big( \hat{\phi}_{\bm{p}} + \hat{\phi}_{-\bm{p}}^{\dagger} \big) - gn \bigg] \nonumber \\
&&{}+ \frac{\mathcal{I}N}{2} \sum_{\bm{p},\bm{p}'}{}^{'} \sum_{\bm{k},\bm{k}'\in\mathcal{K}_{\rm C}} \left[ \big( \hat{\phi}_{\bm{p}} \delta_{-\bm{p}\bm{k}}^{(d)} + \hat{\phi}_{\bm{p}}^{\dagger} \delta_{\bm{p}\bm{k}}^{(d)} \big) \tilde{v}_{\bm{k},\bm{k}'} \right. \nonumber \\
&&\left.{} \times \tilde{f}_{\bm{\xi}}(\bm{0},\bm{0}) \big( \hat{\phi}_{\bm{p}'} \delta_{-\bm{p}'\bm{k}'}^{(d)} + \hat{\phi}_{\bm{p}'}^{\dagger} \delta_{\bm{p}'\bm{k}'}^{(d)} \big) \right] \, ,
\label{H2}
\end{eqnarray}
where in the double sum we used the approximation of Eq.\ \eqref{approxKroneckerDelta}.
We introduce the quasi-position operator $\hat{x}_{\bm{p}} = \sqrt{M/\bm{p}^2} ( \hat{\phi}_{\bm{p}} + \hat{\phi}_{-\bm{p}}^{\dagger})$ and the quasi-momentum operator $\hat{y}_{\bm{p}} = -i \sqrt{\bm{p}^2/4M} (\hat{\phi}_{-\bm{p}} - \hat{\phi}_{\bm{p}}^{\dagger})$, which satisfy $\hat{x}_{\bm{p}}^{\dagger} = \hat{x}_{-\bm{p}}$, $\hat{y}_{\bm{p}}^{\dagger} = \hat{y}_{-\bm{p}}$, and $[\hat{x}_{\bm{p}},\hat{y}_{\bm{p}'} ] = i \delta_{\bm{p}\bm{p}'}^{(d)}$ \cite{bogoliubov2003tommasini}. Then, Eq.\ (\ref{H2}) reduces to the expression
\begin{eqnarray}
\label{halfwayDone}
\hspace*{-5mm}\hat{H}_2 &=& \frac{1}{2} \sum_{\bm{p}}{}^{'} \left( \hat{y}_{\bm{p}}^{\dagger} \hat{y}_{\bm{p}} + \omega_{\bm{p}}^2 \hat{x}_{\bm{p}}^{\dagger} \hat{x}_{\bm{p}} - \frac{\bm{p}^2}{2M} - gn \right) \nonumber \\
&&+ \frac{\mathcal{I}N}{2} \tilde{f}_{\bm{\xi}}(\bm{0},\bm{0}) \sum_{\bm{k},\bm{k}' \in \mathcal{K}_{\rm C}} \tilde{v}_{\bm{k},-\bm{k}'} \:\!\frac{|\bm{k}||\bm{k}'|}{M} \,\hat{x}_{\bm{k}}^{\dagger} \hat{x}_{\bm{k}'} \, .
\end{eqnarray}
From the second line, we see that the long-range interaction couples exclusively the quasi-position operators $\hat{x}_{\bm{k}}$ for wave vectors $\bm{k} \in \mathcal{K}_{\rm C}$.
Note that this type of coupling depends on the choice of $v(\bm{r},\bm{r}')$ as an even and symmetric function.
The modes $\bm{p} \notin \mathcal{K}_{\rm C}$, that do not couple to the cavity, follow the familiar Bogoliubov dispersion 
\begin{align}
\omega_{\bm{p}} = \sqrt{\frac{\bm{p}^2}{2M} \left( \frac{\bm{p}^2}{2M} + 2gn \right)} \, .
\label{sWaveDispersion}
\end{align}
Thus, it remains to determine the respective dispersion for the long-range coupled modes $\mathcal{K}_{\rm C} = \lbrace \bm{k}_1, \dots, \bm{k}_{\tilde{d}} \rbrace$, where we denote the number of these modes by $\tilde{d} = |\mathcal{K}_{\rm C}|$.
To this end, we define $\hat{\vec{x}} = \begin{pmatrix} \hat{x}_1 &\dots &\hat{x}_{\tilde{d}} \end{pmatrix}{\!}^T$ and analogously $\hat{\vec{y}}$ as well as the effective coupling
\begin{align}
\tilde{v}_{ij} = \tilde{v}_{\bm{k}_i,-\bm{k}_j} \frac{|\bm{k}_i||\bm{k}_j|}{M} \, ,
\end{align}
between the modes $i$ and $j$.
Hence, the Hamiltonian of Eq.\ (\ref{halfwayDone}) can be rewritten as
\begin{eqnarray}
\label{h-fluc}
\hat{H}_2 &=& \frac{1}{2} \sum_{\bm{p}\notin\mathcal{K}_{\rm C}}{\!\!}^{'} \left( \hat{y}_{\bm{p}}^{\dagger} \hat{y}_{\bm{p}} + \omega_{\bm{p}}^2 \hat{x}_{\bm{p}}^{\dagger} \hat{x}_{\bm{p}} - \frac{\bm{p}^2}{2M} - gn \right) \\
&&{}+ \frac{1}{2} \left[\;\! \hat{\vec{y}}^{\,\dagger} \mathbb{I}_{\tilde{d}\times\tilde{d}} {\;\!}\hat{\vec{y}} + \hat{\vec{x}}^{\,\dagger} \underline{h} {\:\!}\hat{\vec{x}} - \sum_{\bm{k}\in\mathcal{K}_{\rm C}} \left( \frac{\bm{k}^2}{2M} + gn \right) \right] \, . \nonumber
\end{eqnarray}
Finding the remaining eigenmodes $\Omega_{\bm{k}}$ with  $\bm{k} \in \mathcal{K}_{\rm C}$ thus relies on the diagonalization of the real symmetric matrix
\begin{align}
\underline{h} = \text{diag}\big( \omega_1^2,\dots,\omega_{\tilde{d}}^2\big) + \mathcal{I}N\tilde{f}_{\bm{\xi}}(\bm{0},\bm{0}){\;\!} \underline{\tilde{v}} \, ,
\end{align}
where $\omega_i = \omega_{\bm{k}_i}$.
Note that the symmetry of the matrix $\underline{\tilde{v}}$ follows directly from Eq.\ (\ref{v-FT-properties}).
The Fourier transform of the envelope $\tilde{f}_{\bm{\xi}}(\bm{0},\bm{0})$, that carries the dependence on the spatial extent of the system, appears as a mere prefactor to the interaction matrix $\underline{\tilde{v}}$.
In conclusion, the zero-point energy of the quantum fluctuations reads 
\begin{eqnarray}
E_{\rm qf} &=& \frac{1}{2} \sum_{\bm{p}\notin\mathcal{K}_{\rm C}}{\!\!}^{'} \left( \omega_{\bm{p}} - \frac{\bm{p}^2}{2M} - gn \right) \nonumber \\
&&{}+ \frac{1}{2} \sum_{\bm{k}\in\mathcal{K}_{\rm C}} \left( \Omega_{\bm{k}} - \frac{\bm{k}^2}{2M} - gn \right) \, .
\label{quantumCorrection}
\end{eqnarray}    

\subsection{Discussion\label{discussion}}

By implementing the Bogoliubov transformation, we have naturally separated the modes into those unaffected and those affected by the long-range interaction.
Similarly, the energy correction due to quantum fluctuations $E_{\rm qf} = E_{\rm qf,A} + E_{\rm qf,C}$ contains the term $E_{\rm qf,A}$, which is exclusively due to the atomic contact interaction, and the correction $E_{\rm qf,C}$, which occurs only in the presence of the cavity-induced long-range interaction.
To this end, we complete the sum in the first line of Eq.\ (\ref{quantumCorrection}) by extracting the respective terms from the second line and obtain
\begin{eqnarray}
E_{\rm qf,A} &=& \frac{1}{2} \sum_{\bm{p}}{}^{'} \left( \omega_{\bm{p}} - \frac{\bm{p}^2}{2M} - gn \right) \label{scatteringCorrection} \, , \\
E_{\rm qf,C} &=& \frac{1}{2} \sum_{\bm{k}\in\mathcal{K}_{\rm C}} \left(\Omega_{\bm{k}} - \omega_{\bm{k}} \right) 
\label{longRangeCorrection} \, .
\end{eqnarray}
The atomic fluctuation correction $E_{\rm qf,A}$ due to the contact interaction can be evaluated in the continuum limit.
With the proper regularization for the chosen system dimension \cite{Fetter}, this yields in one and three dimensions, respectively, \cite{petrov2015quantum,petrov2016ultradilute}
\begin{align}
E_{\rm qf,A}^{\rm 1D} = - \frac{2L\sqrt{M}}{3\pi} (gn)^{3/2} \, ,\\
E_{\rm qf,A}^{\rm 3D} = \frac{8VM^{3/2}}{15\pi^2} (gn)^{5/2} \, .
\end{align}
For a dilute Bose gas, the quantum correction due to the contact interaction $g$ is of subleading order compared to the atomic mean-field contribution $E_{\rm mf,A}$, regardless of the underlying spatial dimension $d$.
Therefore, without loss of generality, we can limit our discussion to systems where $E_{\rm mf,A} \gg E_{\rm qf,A}$ holds. Thus, from now on we will neglect the quantum fluctuation correction $E_{\rm qf,A}$ due to the contact interaction.
Furthermore, we note that the cavity-induced quantum fluctuation energy correction $E_{\rm qf,C}$ naturally depends on the spatial extent of the system via $\tilde{f}_{\bm{\xi}}(\bm{0},\bm{0})$, which appears within at least some of the eigenmode frequencies $\Omega_{\bm{k}}$. 
This leads to the fundamental conclusion that the spatial average of the envelope $f_{\bm{\xi}}(\bm{r},\bm{r}')$ crucially determines how the quantum fluctuation correction of the long-range interaction depends on the system extension.
In the following, we discuss two generic cases in which it is straightforward to find analytically the eigenmodes involving the long-range interaction.

\begin{figure}
\centering
\includegraphics[scale=.4]{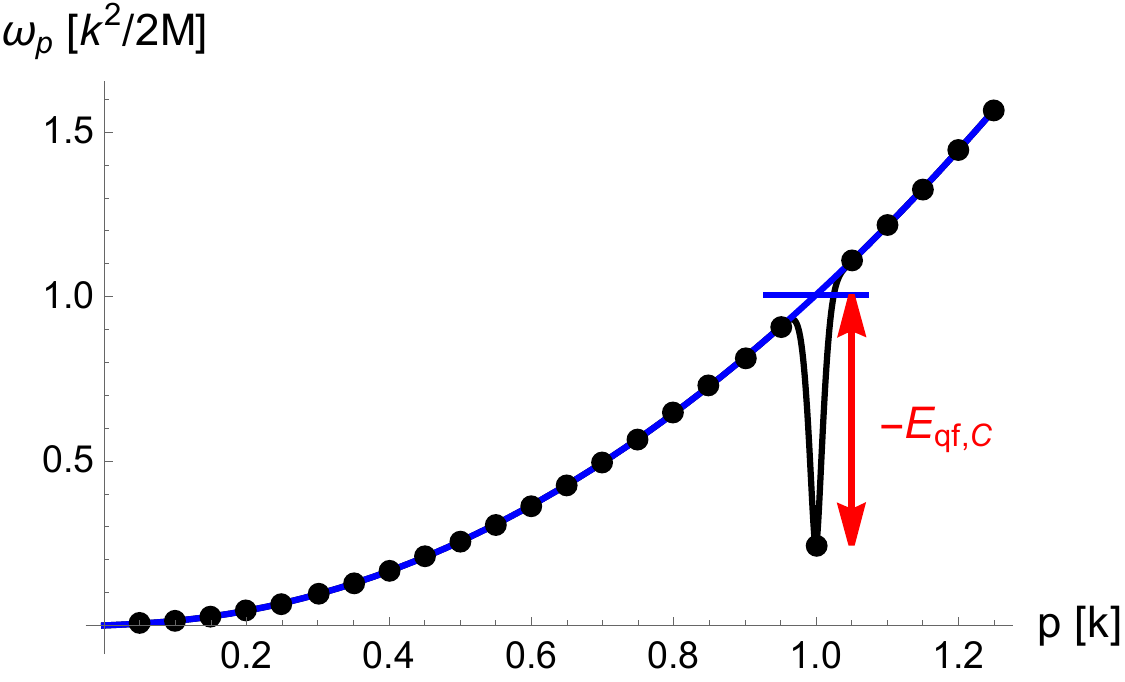}
\caption{Dispersion relation Eq.\ (\ref{translationInvariantDispresion}) for the simple one-dimensional example of $\mathcal{K}_{\rm C} = \lbrace -k,k \rbrace$. Dots mark the discrete modes of the finite system with the distinct roton at $p=k$. The black line indicates the continuum limit of the discrete modes indicated by the dots. The blue solid line shows the continuum of dispersion without the long-range interaction, $\mathcal{I}=0$. The horizontal blue line marks the roton mode value for $\mathcal{I}=0$. Consequently, the difference between the actual roton value and the blue horizontal line shown by the red arrow represents the roton contribution to the long-range induced quantum correction $E_{\rm qf,C}$.}
\label{dispersion}
\end{figure}

First, we consider a long-range interaction that is translationally invariant and thus momentum conserving.
In principle, both its components $v(\bm{r},\bm{r}')$ and $f_{\bm{\xi}}(\bm{r},\bm{r}')$ must be translationally invariant.
However, within the conditions imposed on the envelope that allow us to state Eq.\ (\ref{approxKroneckerDelta}) and derive Eq.\ (\ref{h-fluc}), the envelope contributes only a prefactor $\tilde{f}_{\bm{\xi}}(\bm{0},\bm{0})$.
Therefore, for a momentum conserving interaction it is sufficient that the matrix $\underline{\tilde{v}}$ is diagonal, i.e., that the periodic function $v(\bm{r},\bm{r}')$ is translationally invariant.
Then, the second line in Eq.\ (\ref{halfwayDone}) has the consequence that only those modes $\bm{k}$ and $\bm{k}'$ couple which satisfy the condition $\bm{k} = \bm{k}'$.
Consequently, the fluctuation Hamiltonian $\hat{H}_2$, expressed in the quasi-position and quasi-momentum operators, turns out to be already diagonal, i.e.,
\begin{align}
\hat{H}_2 ={}& \frac{1}{2} \sum_{\bm{p}\notin \mathcal{K}_{\rm C}}{\!\!}^{'} \bigg( \hat{y}_{\bm{p}}^\dag \hat{y}_{\bm{p}} + \omega_{\bm{p}}^2 \hat{x}_{\bm{p}}^\dag \hat{x}_{\bm{p}} - \frac{\bm{p}^2}{2M} - gn \bigg) \nonumber\\
&+ \frac{1}{2} \sum_{\bm{k}\in \mathcal{K}_{\rm C}}\bigg( \hat{y}_{\bm{k}}^\dag \hat{y}_{\bm{k}} + \Omega_{\bm{k}}^2 \hat{x}_{\bm{k}}^\dag \hat{x}_{\bm{k}} - \frac{\bm{k}^2}{2M} - gn \bigg)\, ,
\label{translationInvariantFluctuationHamiltonian}
\end{align}
where the dispersion of the modes affected by the long-range interaction reads
\begin{align}
\Omega_{\bm{k}} = \sqrt{\omega_{\bm{k}}^2 + \frac{\bm{k}^2}{M}\, \mathcal{I}N \tilde{f}_{\bm{\xi}}(\bm{0},\bm{0})\:\! \tilde{v}_{\bm{k},-\bm{k}}} \, .
\label{translationInvariantDispresion}
\end{align}
If the long-range interaction is attractive, i.e., $\mathcal{I} < 0$, these modes give rise to a roton at each $\bm{k} \in \mathcal{K}_{\rm C}$ for which $\tilde{v}_{\bm{k},-\bm{k}}\neq 0$.
Thus, the momenta at which a roton is located in momentum space are selected by the wave vectors reflecting the periodicity of the long-range interaction $v(\bm{r},\bm{r}')$.
We visualize this phenomenon in Fig.~\ref{dispersion} for a simple one-dimensional example with $\mathcal{K}_{\rm C} = \lbrace -k,k \rbrace$.
The corresponding quantum correction of the ground-state energy relative to the case without the long-range interaction is given by Eq.\ (\ref{longRangeCorrection}).
For the roton case, it is indeed negative.
Importantly, the roton depth depends on the system size via the spatial average of the envelope $\tilde{f}_{\bm{\xi}}(\bm{0},\bm{0})$.
This directly implies that the quantum energy correction $E_{\rm qf,C}$ (\ref{longRangeCorrection}) depends on the system size.
Thus, its respective derivative $(\partial E_{\rm qf,C}/\partial V)_N$ is added to the mean-field one and gives rise to the conditions for droplet formation in Eq.\ (\ref{condition1}).

Second, we consider the case where $v(\bm{r},\bm{r}')$ has the property that all coupled modes have the same modulus, i.e., $|\bm{k}| = |\bm{k}'|$ for all $\bm{k},\bm{k}' \in \mathcal{K}_{\rm C}$.
Since the modes $\omega_{\bm{k}}$ in Eq.\ (\ref{sWaveDispersion}) depend only on $\bm{k}^2$, we conclude that $\omega_{\bm{k}} = \omega_{\bm{k}'}$ for all $\bm{k},\bm{k}' \in \mathcal{K}_{\rm C}$.
In addition, we assume $\tilde{v}_{ij} = \tilde{v} \bm{k}^2/M$ for all $\bm{k}_i,\bm{k}_j \in \mathcal{K}_{\rm C}$, i.e., all entries in the interaction matrix $\underline{\tilde{v}}$ are equal. 
In this scenario, it is straightforward to analytically find the eigenmode modified by the long-range interaction in the form
\begin{align}
\Omega = \sqrt{\omega_{\bm{k}}^2 + \frac{\bm{k}^2}{M} \,\mathcal{I}N \tilde{f}_{\bm{\xi}}(\bm{0},\bm{0}){\:\!} \tilde{d}{\:\!} \tilde{v}} \, ,
\label{longRangeEigenmode}
\end{align}
whereas the remaining $(\tilde{d}-1)$ eigenmodes turn out to be degenerate and lie in the dispersion $\omega_{\bm{k}}$, i.e., they are unaffected by the long-range interaction.
Therefore, provided that $\mathcal{I} < 0$, then $\Omega$ is a discrete roton mode that softens at 
\begin{align}
\mathcal{I}_{\rm cr} = - \frac{\bm{k}^2/2M + 2gn}{2N\tilde{f}_{\bm{\xi}}(\bm{0},\bm{0}){\:\!}\tilde{d}{\:\!}\tilde{v}}\, .
\label{generalCriticalValueEx}
\end{align}
Furthermore, it contributes to the zero-point energy via
\begin{align}
E_{\rm qf,C} = \frac{1}{2} \left( \Omega - 
\omega_{\bm{k}} \right) \, , 
\label{longRangeZeroPointMotion}
\end{align}
which is negative due to the roton characteristic.
This quantum fluctuation energy of an individual mode has the peculiar property that it is not extensive but intensive, so it would vanish in a proper thermodynamic limit.
Nevertheless, it is a viable energy contribution for a finite-sized system, which leads to intricate consequences for the effective ground-state energy $E_0 = E_{\rm mf,A} + E_{\rm qf,A} + E_{\rm qf,C}$, as it participates in the competition between the respective energy contributions.
Provided that their interplay is such that the droplet conditions (C1)-(C3) are satisfied, a finite quantum droplet is realized.
We have already argued that $E_{\rm qf,A} \ll E_{\rm mf,A}$ is negligibly small.
Hence, the key aspect must be the competition between the long-range induced quantum correction $E_{\rm qf,C}$ and the contact interaction mean-field contribution $E_{\rm mf,A}$.
We find that a roton having $E_{\rm qf,C} < 0$ with an appropriate dependence on the system size imposed by the choice of the envelope $f_{\bm{\xi}}(\bm{r},\bm{r}')$ provides such a suitable competition with a repulsive mean-field energy $E_{\rm mf,A} > 0$.

Formally, the Fourier transform of the envelope $\tilde{f}_{\bm{\xi}}(\bm{0}, \bm{0})$ is a function of $L_{\nu}/\xi_{\nu}$ and can be expanded around $L_{\nu}/\xi_{\nu} = 0$ in the limit $\xi_{\nu} \to \infty$ to get a qualitative insight into how the shape of the envelope enters the long-range interaction correction in Eqs.\ (\ref{translationInvariantDispresion}) and (\ref{longRangeZeroPointMotion}) through the roton frequency $\Omega_{\bm{k}}$.
Due to the restrictions imposed on the envelope, the atoms effectively see only its spatial average, so that even the expansion to the first non-trivial order in $L_{\nu}/\xi_{\nu}$ gives quite a good quantitative approximation.
This allows us to derive a qualitative effective minimal model analogous to Eq.\ (\ref{minimalModel}) from the long-range quantum correction $E_{\rm qf,C}$.

\section{Factorized Envelope in a Single-Mode Cavity\label{factorizedResults}}
\begin{figure}
\centering
\includegraphics[scale=.3]{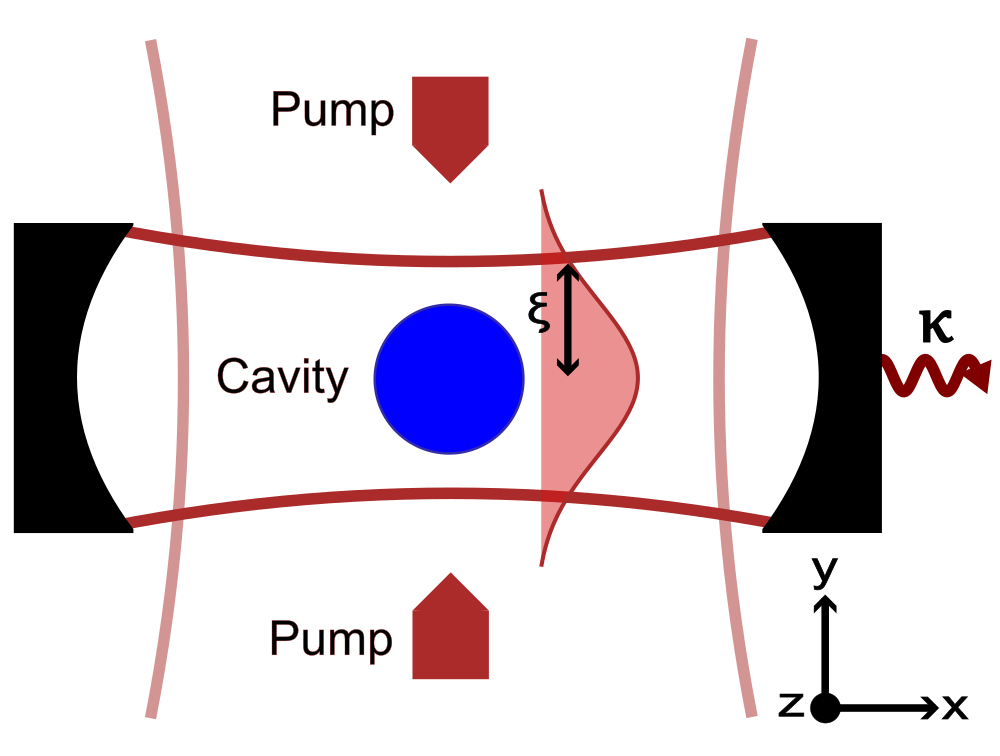}
\caption{Sketch of a cavity BEC setup with a single-mode cavity. The BEC in blue is pumped by a broad beam along both directions of the $y$-axis. The cavity axis is along the $x$-axis with photon loss  through the right mirror with a rate $\kappa$. The cavity mode has a Gaussian transverse profile with waist $\xi$. \label{singleSetup}}
\end{figure}
A factorized long-range interaction is realized, for instance, in the setup of a three-dimensional BEC coupled to a single cavity mode as sketched in Fig.~\ref{singleSetup}.

\subsection{Setup}

The BEC of $N$ two-level atoms is pumped by a transverse beam in a Jaynes-Cummings type coupling \cite{maschler2008ultracold}.
Its theoretical description is performed in a frame rotating at the pump frequency $\omega_{\rm P}$ within the rotating wave approximation \cite{maschler2008ultracold}.
It is also assumed that the pump detuning $\Delta_{\rm A} = \omega_{\rm P} - \omega_{\rm A} < 0$ with respect to the internal atomic transition frequency $\omega_{\rm A}$ is large, so that the excited atomic state can be adiabatically eliminated.
The system is then described by the cavity mode $\hat{a}$, detuned by $\Delta_{\rm C} = \omega_{\rm P} - \omega_{\rm C} < 0$, and
by the bosonic field operator $\hat{\psi}(\bm{r})$ of the atomic ground-state.
The corresponding Hamiltonian of the cavity BEC system \cite{maschler2008ultracold,mivehvar2021cavity} reads
\begin{flalign}
\label{singleCavityFieldHamiltonian}
&\hat{H} = \int_V d^3\bm{r} \left\lbrace  \hat{\psi}^{\dagger}(\bm{r}) \left[ -\frac{\bm{\nabla}^2}{2M} + \frac{h^2(\bm{r})}{\Delta_{\rm A}} +  \frac{\mathcal{G}(\bm{r}) h(\bm{r})}{\Delta_{\rm A}} (\hat{a} + \hat{a}^{\dagger}) \right. \right. & \nonumber\\
& \hspace{7mm} + \left. \left. \frac{\mathcal{G}^2(\bm{r})}{\Delta_{\rm A}} \hat{a}^{\dagger} \hat{a} + \frac{g}{2} \hat{\psi}^{\dagger}(\bm{r}) \hat{\psi}(\bm{r}) \right] \hat{\psi(\bm{r})} \right\rbrace - \Delta_{\rm C} \hat{a}^{\dagger} \hat{a} \, . \hspace{-7mm} & 
\end{flalign}
We assume that the pump beam propagates along the $y$-axis and is so broad that we can neglect its transverse spatial dependence and have only a two-dimensional envelope, which will be introduced below.
In such a case, the pump mode function is simply $h(\bm{r}) = h_0 \cos{(ky)}$, where $h_0$ stands for the pump Rabi frequency, and we can neglect the pump influence on the atomic confinement in the $xz$-plane.
The penultimate term in the first line of Eq.\ \eqref{singleCavityFieldHamiltonian} is due to pump self-interference after back reflection.
To focus on the central features of the system, we will neglect this term, since it can be canceled in the experiment by an additional field along the pump axis.
We assume that the cavity mode is TEM$_{00}$ and denote the Rabi frequency of the coupling to the atoms at the mode's center by $\mathcal{G}_0$.
Thus, the cavity mode function $\mathcal{G}(\bm{r}) = \mathcal{G}_0 \cos{(kx)} \exp{[-(y^2+z^2)/\xi^2]}$ provides a Gaussian envelope of waist $\xi$ transverse to the cavity axis in $x$-direction.
The Rabi frequency $\mathcal{G}_0$ is proportional to the cavity electric field strength and the single atom-cavity coupling strength $U_0$ and is determined by $U_0 = \mathcal{G}_0^2/\Delta_{\rm A}$.
The last term in the first line of Eq.\ (\ref{singleCavityFieldHamiltonian}) contains a linear coupling of the cavity photons to the atoms, which is enhanced by the scattering of the pump light.
This term is crucial as it is responsible for the cavity population and allows the transition from an empty cavity and homogeneous cloud to a superradiant self-organized state via the Dicke quantum phase transition \cite{nagy2010dicke}.
The first term in the second line describes the optomechanical interaction of the cavity field with the atomic cloud.
Finally, the last term in the integral of Eq.\ (\ref{singleCavityFieldHamiltonian}) represents the atom-atom contact interaction described by the pseudopotential strength $g = 4 \pi a_{\rm s} / M$, where $a_{\rm s}$ denotes the $s$-wave scattering length.
In this work, we consider the following hierarchy of parameter values $|\Delta_{\rm A}| \gg |\Delta_{\rm C}| \gg \omega_{\rm R} = k^2/(2M) \gg g n,\:\!|U_0|$, which is congruent with available experimental setups. Furthermore, $\omega_{\rm R}$ is the recoil energy and
we assume that the atoms are contained in the box of size $L = V^{1/3} < \xi$, which is enclosed by the cavity mode.

\subsection{Effective model}

Cavity BEC experiments have inherent photon losses that allow a non-destructive observation of the system by measuring the outcoupled light \cite{mivehvar2021cavity}. Also the quantum fluctuations can be detected by non-destructively measuring the quantum fluctuations of the escaping photons \cite{cavityphotonstatistics}. As usual, the losses can be modeled by white noise fluctuations $\hat\Xi(t)$ obeying $\langle \hat\Xi(t)\rangle =0$ and $\langle \hat\Xi(t)\hat\Xi^{\dag}(t')\rangle = 2 \kappa \delta(t-t')$, where $\kappa$ stands for the cavity loss rate.
The cavity mode dynamics is then described in the Heisenberg picture by the quantum Langevin equation 
\cite{nagy2011critical}
\begin{align}
i \:\!\frac{d\hat{a}}{dt} = \big[\hat{a}, \hat{H}\big] - i \kappa \hat{a} + i \:\!\hat\Xi \, .
\end{align}
In the presence of the envelope, it has the form
\begin{align}
i \:\!\frac{d\hat{a}}{dt} = \big( -\Delta_{\rm C} - i \kappa + U_0\:\!\hat{S}' \big) \hat{a} + \frac{\mathcal{G}_0 h_0}{\Delta_{\rm A}}\:\! \hat{S} + i \:\!\hat{\Xi} \, ,
\end{align}
where we have introduced
\begin{align}
\hspace*{-2mm}\hat{S} &= \int_V d^3\bm{r} \cos(kx)\cos(ky)\;\! e^{-(y^2+z^2)/\xi^2} \hat{\psi}^{\dagger}(\bm{r}) \hat{\psi}(\bm{r}) \, ,\\
\hspace*{-2mm}\hat{S}'\! &= \int_V d^3\bm{r} \cos^2(kx)\;\! e^{-2(y^2+z^2)/\xi^2} \hat{\psi}^{\dagger}(\bm{r}) \hat{\psi}(\bm{r}) \, .
\end{align}
Due to the large cavity detuning $|\Delta_{\rm C}|$ and the damping $\kappa$, we can determine the cavity field as a steady state in a Born-Oppenheimer approximation \cite{mivehvar2021cavity} in the form
\begin{align}
\hat{a} = \frac{\mathcal{G}_0h_0}{\Delta_{\rm A}(\Delta_{\rm C}+i\kappa-U_0\:\!\hat{S}')} \,\hat{S} \, .
\end{align}
Next, we expand the denominator in powers of $U_0\:\! \hat{S}'\!/(\Delta_C+i\kappa)$, keep only the 
zeroth-order term, and obtain
\begin{align}
\hat{a} = \frac{\mathcal{G}_0h_0}{\Delta_{\rm A}(\Delta_{\rm C}+i\kappa)} \,\hat{S} \, .
\label{eliminatedLightField}
\end{align}
The operator $\hat{S}'$ occurs explicitly only in higher-order terms, which physically correspond to the interaction of three or more particles.
Thus, omitting $\hat{S}'$ is equivalent to restricting the description to two-body interactions only.
This is feasible since we consider the parameter regime where the number of cavity photons is small and $U_0\|\hat{S}'\|/|\Delta_C+i\kappa|\ll 1$.
The remaining ambiguities in the operator ordering are resolved by Jäger et al.\ in Ref.\ \cite{jager2022lindblad}, which ultimately leads to the effective Hamiltonian 
\begin{align}
\hat{H}_{\rm eff} = \hat{H}_{\rm A} + \frac{\mathcal{G}_0 h_0}{2\Delta_{\rm A}}\, (\hat{a}^{\dagger} \hat{S} + \hat{S}^{\dagger} \hat{a})\, .
\end{align}
With this, we obtain, up to the order $\mathcal{O}(U_0^2/(\Delta_{\rm C}^2+\kappa^2))$, the effective Hamiltonian in Eq.\ (\ref{effectiveInteractingBEC})
for the atomic field,  representing the  model for droplet formation in Ref.~\cite{companionLetter}. Here the cavity-induced interaction turns out to be of the form given in Eq.\ (\ref{longRangePotential}).
Combining cavity and pump parameters leads to the interaction strength 
\begin{align}
\mathcal{I} = \frac{2 \mathcal{G}_0^2 h_0^2 \Delta_{\rm C}}{\Delta_{\rm A}^2(\Delta_{\rm C}^2+\kappa^2)} \, .
\label{cavity-I}
\end{align}
Furthermore, the periodic potential reads
\begin{align}
v(\bm{r},\bm{r}') = \cos(kx)\cos(ky)\cos(kx')\cos(ky')
\end{align}
and the envelope is given by
\begin{align}
\label{env}
f_{\xi}^{(2)}(\bm{r},\bm{r}') = e^{-(y^2+z^2)/\xi^2 -(y'^2+z'^2)/\xi^2}\, ,
\end{align}
where the superscript (2) refers to the second-order powers in the exponents.
Thus, we can directly apply the general formalism developed in Sec.\ \ref{genericModel}.
The transformation of the envelope $\tilde{f}_{\xi}^{(2)}(\bm{p},\bm{p}')$ turns out to be given by non-elementary integrals.
Conveniently, for $m\neq 0$ we have the inequality that is graphically illustated in Fig.\ \ref{graphicalVerification} (a) in Appendix \ref{inequalityVerification}
\begin{align}
\bigg|\int_{-\frac{L}{2}}^{+\frac{L}{2}}\! \frac{dx}{L}\:\! e^{i\frac{2\pi m}{L}x} e^{-\frac{x^2}{\xi^2}} \:\!\bigg| < \frac{L^2}{2 \pi^2 m^2 \xi^2} \int_{-\frac{L}{2}}^{+\frac{L}{2}}\! \frac{dx}{L}\:\! e^{-\frac{x^2}{\xi^2}} \, .
\label{singleTransformationEstimate}
\end{align}
Thus, for $\bm{p},\bm{p}' \neq \bm{0}$ we get the estimate
\begin{align}
\hspace{-1mm}\frac{\big|\tilde{f}_{\xi}^{(2)} (\bm{p},\bm{p}')\big|}{\tilde{f}_{\xi}^{(2)} (\bm{0},\bm{0})} < \prod_{\nu=1,2}{}^{\!\!\!'} \frac{L^2}{2 \pi^2 m_\nu^2\xi^2} \times \prod_{\nu=1,2}{}^{\!\!\!'} \frac{L^2}{2 \pi^2 m_\nu'^2\xi^2} \, ,
\end{align}
where the primed products exclude the terms $m_\nu^{} = 0$ and $m_\nu' = 0$. Noticing that $(2\pi^2)^{-1}\approx 0.05$, we conclude that Eq.\ (\ref{approxKroneckerDelta}) is valid as long as $L < \xi$.
The envelope is then taken into account through its $\bm{p}=\bm{p}'=\bm{0}$ Fourier series coefficient (\ref{envelopeFourierSeries})
\begin{align}
\tilde{f}_{\xi}^{(2)}(\bm{0},\bm{0}) = \left[ \frac{\sqrt{\pi} \xi}{L} \,\text{erf}\left( \frac{L}{2\xi} \right) \right]^4 \, .
\label{special-case}
\end{align}
Thus, Eq.~(\ref{longRangeMeanField}) is directly applicable together with Eq.\ (\ref{special-case}). Since the wavenumber $k$ of the light field is non-zero,
the long-range periodic potential $v(\bm{r},\bm{r}')$ only couples the four modes $\mathcal{K}_{\rm C} = \lbrace \begin{pmatrix} \pm k & \pm k & 0 \end{pmatrix}^T \rbrace $ with the same modulus $\sqrt{2}k$ and we have $\tilde{v}_{\bm{k},\bm{k}'} = 1/16$ for all $\bm{k},\bm{k}'\in\mathcal{K}_{\rm C}$.
The conditions of Eq.~(\ref{longRangeEigenmode}) are fulfilled, so the quantum fluctuation correction of the cavity modes is given by Eq.\ (\ref{longRangeZeroPointMotion}), with the roton energy
\begin{eqnarray}
\label{singleCavityRoton}
\Omega = \sqrt{ \omega_{\bm{k}}^2 + \frac{k^2}{M} \frac{\mathcal{I}N}{2} \left[ \frac{\sqrt{\pi}\xi}{L} \, \text{erf}\left( \frac{L}{2\xi} \right) \right]^4 } \, ,
\end{eqnarray}
and the Bogoliubov dispersion
\begin{align}
\label{singleCavityBogoliubov}
\omega_{\bm{k}} = \sqrt{\frac{k^2}{M} \left(\frac{k^2}{M} +2gn\right)} \, .
\end{align}

\subsection{Results}
\begin{figure}
    \centering
    \includegraphics[scale=.35]{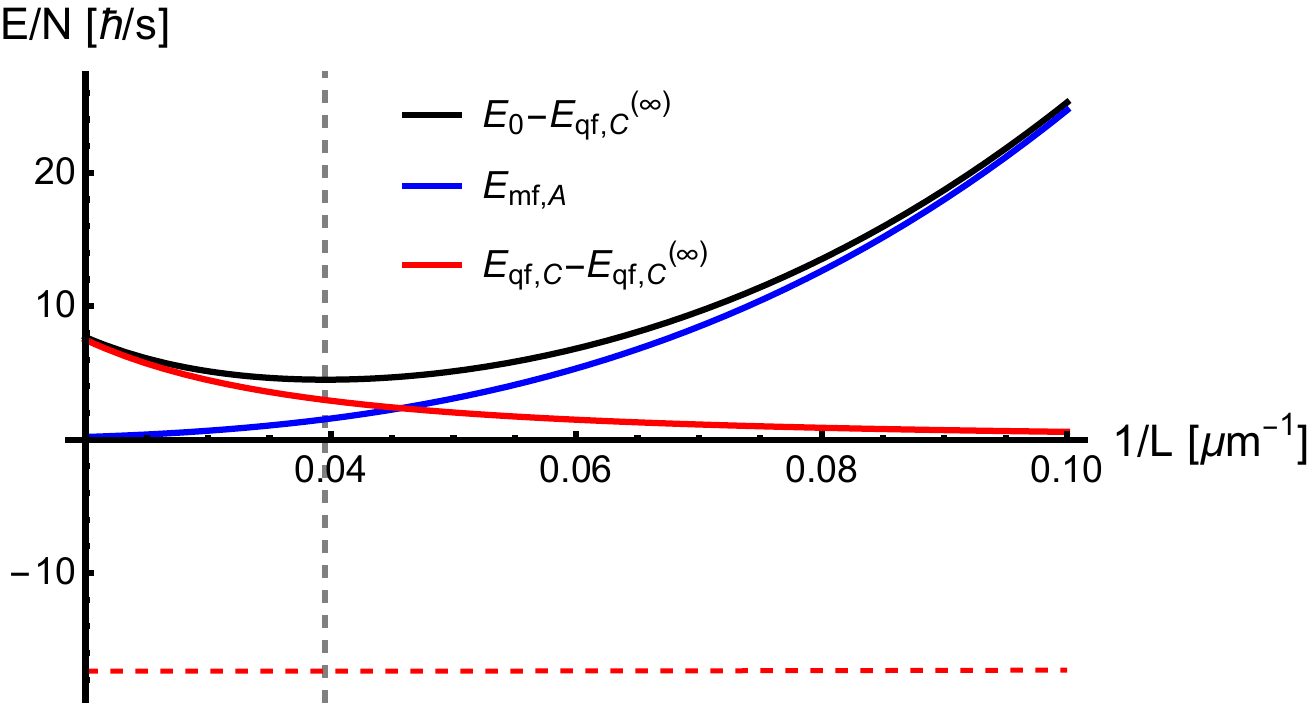}
    \caption{Effective ground-state energy $E_0$ per particle with its mean-field $E_{\rm mf,A}$ and cavity-induced quantum fluctuations contributions $E_{\rm qf,C}$ plotted against the inverse of the length of the atomic cloud $L$ \cite{companionLetter}. The energy correction of an infinite-range cavity $E_{\rm qf,C}^{(\infty)}$, i.e., $L/\xi \to 0$, is subtracted in order to bring together the curves and is plotted for the sake of comparison as a constant red dashed line. An 
    energy minimum is realized marked by the vertical gray dashed line indicating an equilibrium droplet size at $1/L_0$.
    Other parameters are $N=10^3$,  $\mathcal{I} = -85$ Hz, $\xi=50\, \mu$m, $a_{\rm s} = 100\, a_0$, and $M=87\,u$. \label{energyPerParticle}}
\end{figure}
We analyze the effective energy  $E_0 = E_{\rm mf,A} + E_{\text{qf,C}}$ in accordance with the aforementioned criteria for the existence of quantum droplets (C1)--(C3) \cite{companionLetter}.
Primarily we require that for a fixed number $N$ of atoms the system realizes an energy minimum, according to (C1) and (C2), which results in an equilibrium volume $V_0$. 
To this end, we discriminate between the competing contributions of the energy $E_0$.
For the repulsive contact interaction of a stable BEC, consisting of, e.g., $^{87}$Rb atoms, the mean-field energy $E_{\rm mf,A}=gN^2/(2V)$ is positive. By itself it does not have a minimum at a finite volume.  
Only due to the competition with the cavity quantum fluctuations Eq.\ (\ref{longRangeZeroPointMotion}), together with Eqs.\ 
(\ref{singleCavityRoton}) and (\ref{singleCavityBogoliubov}), an energetic minimum can occur
as shown in Fig.~\ref{energyPerParticle}.
Imposing a negative cavity detuning $\Delta_{\rm C} < 0$ leads to a negative cavity parameter  in Eq.\ (\ref{cavity-I}), i.e., $\mathcal{I}< 0$, implying a negative cavity quantum correction $E_{\text{qf,C}}<0$ based on its roton characteristics.
As our theory is based on a homogeneous mean-field phase, we must stay below the self-organizing superradiant Dicke phase transition. 
The latter occurs, when the roton-mode in Eq.\ (\ref{singleCavityRoton}) becomes soft, i.e., when $\Omega=0$, from which we determine the critical value
\begin{align}
\mathcal{I}_{\rm cr} = -\, \frac{2 \left( k^2/M + 2gn\right)}{N \left[ \sqrt{\pi} \xi \erf \left( L/2\xi \right)/L \right]^4} \, .
\label{DickeCriticalPoint}
\end{align}
In the limit of $\xi \to \infty$, where the square bracket in the denominator approaches $1$, this agrees with the well-established result for cavities with infinite-range interactions \cite{nagy2011critical}.
The fact that we are dealing with the zero-point motion of an individual roton mode carries fascinating implications for the quantum energy correction $E_{\rm qf} \approx E_{\text{qf,C}}$. In the thermodynamic limit one takes $N,L,\xi \to \infty$, while the atom density $N/V$ and the ratio $L/\xi$ of the system length $L$ and the pump waist $\xi$ remain constant.
Furthermore, the coupling of an individual atom to the cavity vanishes $\mathcal{G}_0 \to 0$ so that $\mathcal{I} V$ remains constant \cite{piazza2013bose}.
Thus, the energy contribution of the roton is then intensive, rendering the quantum droplet formation a finite-size effect.
This has profound consequences as the largest possible cavity energy correction at $\Omega = 0$ is eventually overwhelmed by any extensive energy term in the limit of large $N$.

\subsection{Analytic Approximations}

To get a deeper insight into the cavity energy, we perform some analytical approximations to the energy contribution of the quantum fluctuations $E_{\text{qf,C}}$.
For this purpose we recall that the requirement of a homogeneous mean field implies the restriction $L/\xi < 1$.
Physically, this means that the atomic system effectively perceives only the center of the envelope.
This justifies the subsequent expansion of the envelope in Eq.\ (\ref{env}) up to the second order with respect to $L/\xi$, i.e., 
\begin{align}
\hspace*{-3mm}f_{\xi}^{(2)}(\bm{r},\bm{r}') = 1 - \frac{y^2}{\xi^2} - \frac{z^2}{\xi^2} - \frac{y'^2}{\xi^2} - \frac{z'^2}{\xi^2} + \mathcal{O} \left( \frac{L^4}{\xi^4} \right) \, .
\end{align}
With $L$ denoting the system extension in each dimension, i.e., $|y|,|z|,|y'|,|z'| < L/2$, already the second-order expansion turns out to be quite accurate for $L/ \xi < 1$. Applying this approximation to the roton mode in Eq.\ (\ref{singleCavityRoton}) yields
\begin{align}
\Omega \approx \sqrt{ \omega_{\bm{k}}^2 + \frac{k^2}{M} \frac{\mathcal{I}N}{2} } - \frac{\mathcal{I}N}{12 \sqrt{1 + (4gn + \mathcal{I}N)M/2k^2}} \,\frac{L^2}{\xi^2} \, ,
\end{align}
up to the term $\mathcal{O}(L^4/\xi^4)$.
In the following we denote the first zeroth-order term in $L/\xi$ as $\Omega^{(\infty)}$, since it is the roton mode of the cavity with infinite-range interactions, i.e., for $\xi \to \infty$. This demonstrates explicitly that we can recover the results known for the infinite-range interaction at any point during the calculation. The subsequent term  of order $L^2/\xi^2$ then carries the characteristic information of the envelope $f_{\xi}^{(2)}(\bm{r},\bm{r}')$, which is imprinted on the cavity-induced roton mode $\Omega$.
Due to the quadratic dependence on $L$, we cast it in the form of a harmonic oscillator potential $DL^2/2$ with the effective spring constant 
\begin{eqnarray}
D = \frac{-\mathcal{I}N}{12\xi^2\sqrt{1 +(4gn+\mathcal{I}N)M/2k^2}} \, .
\label{harmonicConfinement}
\end{eqnarray}
However, we have to take into account that Eq.\ (\ref{harmonicConfinement}) still depends for fixed particle number $N$ on the system volume $V$ via the $s$-wave scattering term $gn$.
Below the Dicke phase transition $\mathcal{I}_{\rm cr}$, where the radicant of the square root in the denominator of Eq.\ (\ref{harmonicConfinement}) approaches zero, we can use the fact that $gnM/k^2 \ll 1$. This amounts to discarding terms, which involve the $s$-wave scattering, in both the roton $\Omega$ and its ground-state energy $E_{\rm qf,C}$,
leading to
\begin{align}
E_{\rm qf,C}^{(\infty)} \approx \frac{1}{2}\Bigg[\sqrt{\frac{k^2}{M}\bigg(\frac{k^2}{M} + \frac{\mathcal{I}N}{2}\bigg)} - \frac{k^2}{M}\Bigg] \, ,
\end{align}
alongside with $D \approx {}-{}\mathcal{I}N/(12\xi^2\sqrt{1+\mathcal{I}NM/2k^2})$. 
The underlying effective potential then reads 
\begin{align}
E_0(N,V) = E_{\rm qf,C}^{(\infty)} + \frac{gN^2}{2V} + \frac{D}{2} \,V^{2/3} \, .
\label{effectiveModel}
\end{align}
We can directly relate the respective terms in this expression to the parameters of the minimal model Eq.\ (\ref{minimalModel}).
The infinite-range interaction cavity correction $E_{\rm qf,C}^{(\infty)}$ is within the  approximations independent of the atomic system size and can, thus, be understood as a constant energy shift as illustrated in Fig.~\ref{energyPerParticle}. 
The second term in Eq.\ (\ref{effectiveModel})
represents the mean-field contribution to the effective energy and is linear in $1/V$, thus the identification with Eq.\ (\ref{minimalModel}) leads to $\alpha = gN^2/2$.
Restricting ourselves to a stable BEC implies $\alpha >0$ due to the positive $s$-wave scattering interaction strength $g>0$.
The third term in Eq.\ (\ref{effectiveModel}) is proportional to $V^{2/3}$, which corresponds in the minimal model Eq.\ (\ref{minimalModel}) to the term
$V^{-1-\gamma}$ with $\gamma = -5/3$. From the corresponding prefactor we find $\beta = D/2$.
For a roton mode we have to choose $\mathcal{I}<0$ and, therefore, we conclude $\beta > 0$.
Summing up the model parameters, the competition of a repulsive $s$-wave scattering and and a cavity-induced roton mode results in $\alpha>0$, $\beta>0$, and $\gamma<-1$ such that we the cavity-induced quantum droplets correspond to droplet class (D3) of Eq.~(\ref{dropletClasses}).
The size of these droplets, i.e., the equilibrium system volume $V_0$, can be determined analytically within the minimal model Eq.\ (\ref{effectiveModel}) by inserting the parameter values into the general solution from Eq.\ (\ref{condition1}), yielding
\begin{align}
V_0 = \left(- \frac{18\xi^2gN}{\mathcal{I}}
\,  \sqrt{1+\frac{\mathcal{I}NM}{2k^2}}\,\right)^{3/5}\, .
\label{dropletSolution}
\end{align}
In this way, we have determined 
\begin{figure}
\includegraphics[scale=.4]{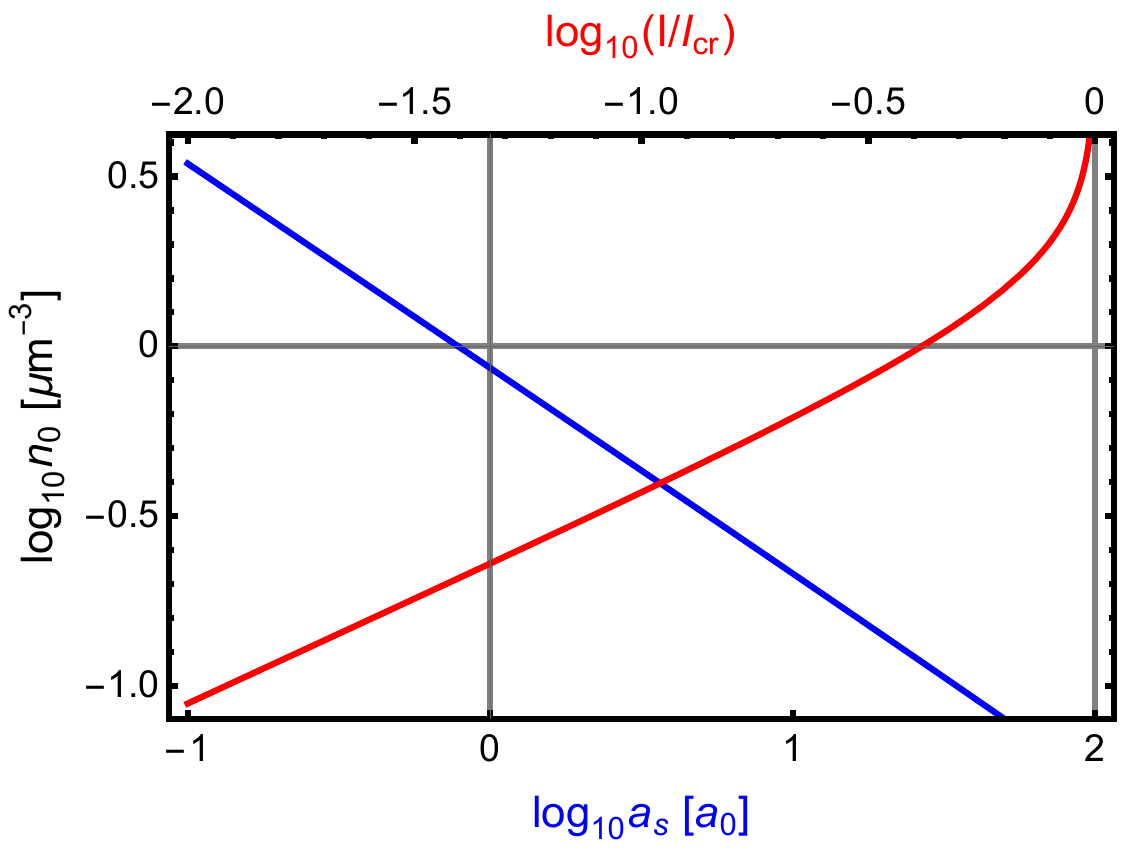}
\caption{Dependence of the droplet density $n_0$ on the interaction parameters, i.e., on the  cavity interaction strength $\mathcal{I}$ (red, top scale) and the s-wave scattering length $a_{\rm s}$ (blue, bottom scale). We set  $a_{\rm s} = 0.1 \, a_0$ and $\mathcal{I} = 0.95 \, \mathcal{I}_{\rm cr}$, respectively.
The lower bound of the $y$-axis is given by the self-consistency constraint $V_0 < \xi^3$.
Parameters are $N=10^4$, $\xi = 50 \, \mu$m, and $M = 87 \, u$. \label{setupParameters}}
\end{figure}
how the droplet size varies with the respective system parameters in leading order.
In terms of the contact interaction strength $g$ we find $V_0 \sim g^{3/5} \sim a_{\rm s}^{3/5}$.
In addition, the dependence on the envelope width, which in the present realization is given by the cavity waist $\xi$, reads $V_0 \sim \xi^{6/5}$.
Here, we have to keep in mind that the theory is constrained to $L/\xi < 1$, and, therefore, the droplet volume must be restricted to $V_0 < \xi^3$ for self-consistency.
Below the Dicke phase transition, which occurs for the vanishing of the radicant of the square root in Eq.\ (\ref{dropletSolution}), we obtain the dependence $V_0 \sim N^{3/5}$ on the atom number. The tunability with respect to the cavity-induced interaction strength $\mathcal{I}<0$ is given by $V_0 \sim |\mathcal{I}|^{-3/5}$.

In Fig.~\ref{setupParameters}, we show how the droplet density $n_0$ depends on the interaction parameters $g$ and $\mathcal{I}$.
Note that the droplet density deviates from the expectation $n_0 \sim |\mathcal{I}|^{3/5}$ in Fig.~\ref{setupParameters} because the cavity interaction strength $\mathcal{I}$ approaches the Dicke critical point $\mathcal{I}_{\rm cr}$ when the square root in Eq.\ (\ref{dropletSolution}) takes significant effect.
Furthermore, the theory visualized in Fig.~\ref{energyPerParticle} predicts that the densities of the cavity-induced quantum droplets are orders of magnitude more dilute than both the observed quantum droplets in dipolar Bose gases or Bose-Bose mixtures \cite{ferrier2016observation, cabrera2018quantum,semeghini2018self,skov2021observation} and the BECs commonly prepared in experiments \cite{mottl2012roton,baumann2010dicke,klinder2015observation}.
This is mainly due to the fact that here the mean-field contact interaction competes with the quantum fluctuation correction, while the standard droplet realizations use Feshbach resonances to almost completely suppress the mean-field contribution.
The relation $V_0 \sim g^{3/5}$ visualized in Fig.~\ref{setupParameters} indicates that a similar suppression of the mean-field would result in an increase of the droplet density by one to two orders of magnitude.
\begin{figure}
    \centering
    \includegraphics[scale=.35]{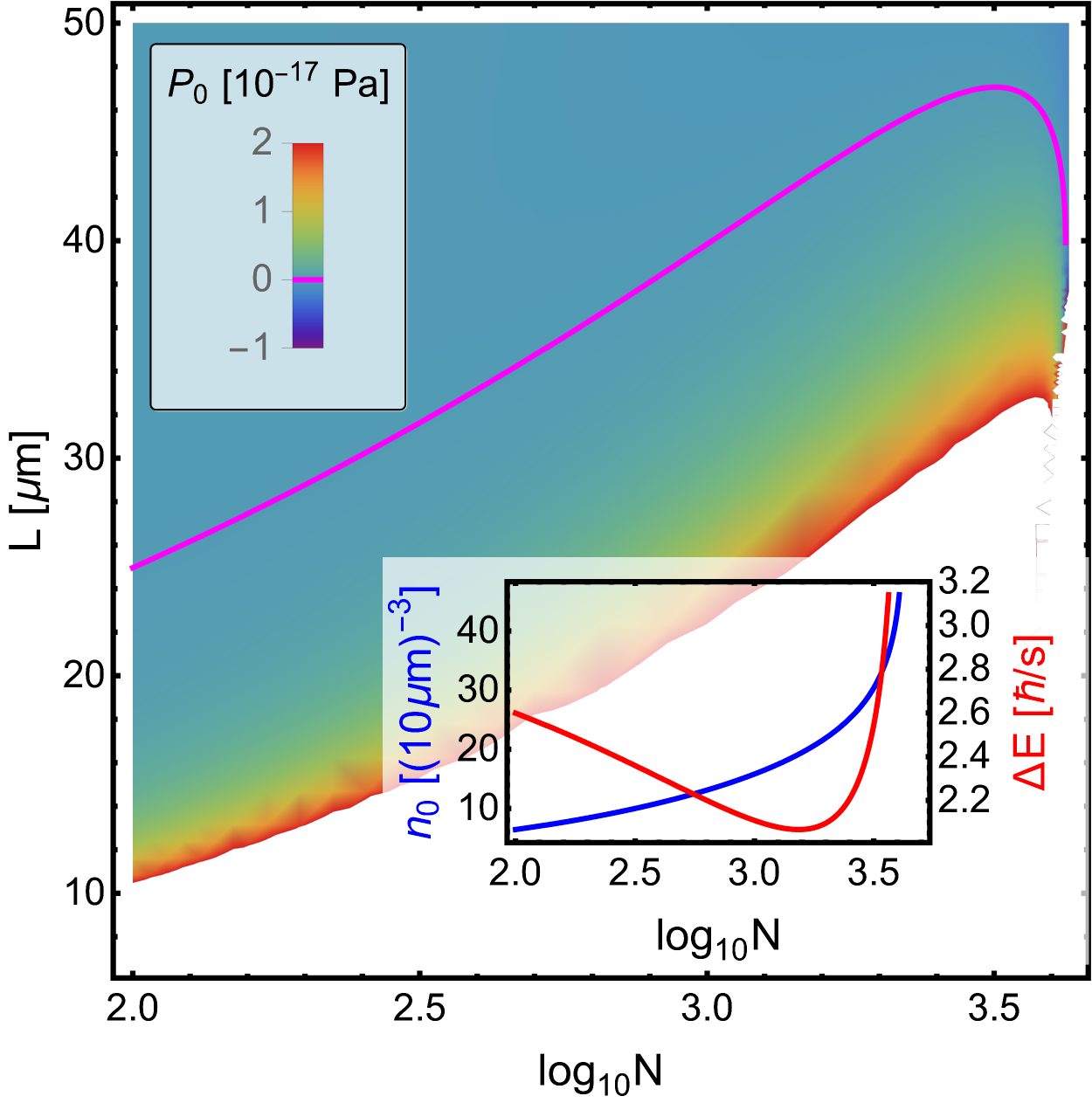}
    \caption{System pressure $P_0$ as a function of the number $N$ of atoms and the system extension $L$ for the cavity interaction strength $\mathcal{I} = -25$ Hz. The magenta line marks zero pressure $P_0 = 0$ corresponding to the droplet solution. The inset shows the density and the single-particle energy difference $\Delta E = E_0(N-1) - E_0(N)$ in blue and red, respectively, along the magenta line shown in the main plot. We note that the roton mode grows softer with increasing number of atoms, therefore the cavity energy correction grows stronger. Remaining parameters are the same as in Fig.~\ref{energyPerParticle}.  \label{pressureDiagram}}
\end{figure}

\subsection{Thermodynamic Properties}

In this section, we discuss the cavity-induced droplets from the point of view of  statistical mechanics.
The first droplet condition (C1) in Eq.\ (\ref{condition1}) translates to a thermodynamic system with zero total pressure, i.e., $P_0 = - (\partial E_0/\partial V)_N = 0$. On the one hand,
the mean-field contact interaction leads to the positive pressure $P_{\rm mf} = gn^2/2$.
On the other hand, the roton contributes the negative pressure $P_{\rm qf,C} = - D/(3L)$, which competes with the mean-field pressure in order to realize an energetic minimum.
The total pressure of the system is plotted in Fig.~\ref{pressureDiagram} as a function of the number $N$ of atoms  and the system size $L$. 
The droplet solution is characterized by the equilibrium system size $L_0$, which corresponds to zero pressure $P_0 = 0$ for each number $N$ of atoms along the magenta line.
The pressure is negative above this line and positive below it.
The latter corresponds to a positive compressibility $K(P_0=0) = -V(\partial P_0/\partial V)_N|_{V=V_0} > 0$, which is the thermodynamic counterpart of the droplet condition (C2).
It obeys $K(P_0=0)/n_0 < gn_0$ implying that also the corresponding speed of sound $c_s = \sqrt{K(P_0=0)/Mn_0}$ is modified accordingly by the quantum fluctuation correction.

The inset of Fig.~\ref{pressureDiagram} visualizes the equilibrium density $n_0$ that realizes the zero pressure condition (C1).
It increases monotonically in accordance with the analytical prediction $n_0 = N/V_0 \sim N^{2/5}$ until it diverges near a critical number of atoms of roughly $N_{\rm cr} \approx 4000$.
This divergence occurs because the critical long-range interaction strength $\mathcal{I}_{\rm cr}$, which is required for the occurrence of the self-organizing Dicke phase transition, decreases with increasing $N$ as follows from Eq.\ (\ref{DickeCriticalPoint}). In more physical terms this can also be understood as follows.
Once the roton goes soft, i.e., the radicant of the square root in Eq.\ (\ref{dropletSolution}) approaches zero, the equilibrium system size $V_0$ becomes arbitrarily small.
This leads to a divergence of the droplet density $n_0$, which accompanies the divergence of the quantum fluctuations close to the Dicke phase transition.
Note that the latter also implies a divergence of the quantum depletion that has not yet been taken into account in the Bogoliubov treatment presented above. 
In Ref.\ \cite{companionLetter} we present the pressure diagram of Fig.~\ref{pressureDiagram} by adjusting the cavity-induced interaction strength $\mathcal{I}$
for each atom number $N$ such that it has a constant value $\mathcal{I} = 0.95 \,\mathcal{I}_{\rm cr}$ relative to the Dicke critical point.

Next, we examine the evaporation condition (C3) for the parameters of the  model in Eq.\ (\ref{effectiveModel}), where $\alpha = gN^2/2$ and $\beta = D/2$.
For a finite system we have to check explicitly that the single-particle energy difference $\Delta E = [E_0(N-1)-E_0(N)]|_{P_0=0}$ remains positive, as displayed in the inset of Fig.~\ref{pressureDiagram}.
Analytically, we study its thermodynamic counterpart $(\partial E_0/\partial N)_{V_0} < 0$, which is the negative chemical potential.
The mean-field term $\alpha/V$ in $E_0$ contributes to the chemical potential $\mu_{\rm mf,A}|_{V_0} = gn_0$, which is positive for a repulsive contact interaction.
In addition, we also have to take into account the system size dependent quantum fluctuation correction term proportional to $\beta$, yielding
\begin{align}
\hspace*{-5mm}\left(\frac{\partial DV^{2/3}}{2\partial N} \right)_{V = V_0} =& {}-{} \frac{\mathcal{I}V_0^{2/3}}{24\xi^2 \sqrt{1+\mathcal{I}NM/2k^2}} \nonumber \\
&{}+{} \frac{\mathcal{I}^2NV_0^{2/3}M/2k^2}{48 \xi^2 \left[ 1+ \mathcal{I}NM/2k^2 \right]^{3/2}} \, .
\label{quantumBetaChemicalPotential}
\end{align}
We observe for a roton $\mathcal{I} < 0$ that both terms are positive.
The negative competing part in the chemical potential is, in fact, provided by the infinite-range interaction term
\begin{align}
\label{inf}
\mu_{\rm qf,C}^{(\infty)} = \frac{\partial E_{\rm qf,C}^{(\infty)}}{\partial N} = \frac{\mathcal{I}}{8\sqrt{1+\mathcal{I}NM/2k^2}} \, .
\end{align}
Using Eq.\ (\ref{dropletSolution}) it becomes apparent that $\mu_{\rm qf,C}^{(\infty)}$ compensates the mean-field chemical potential.
Concluding from Eq.\ (\ref{DickeCriticalPoint}) that $\mathcal{I}M/2k^2 \lesssim 1$ and that the system size is limited by the cavity waist according to $V_0^{2/3}/\xi^2 < 1$, we find that the chemical potential contribution stemming from the infinite-range interaction term in Eq.\ (\ref{inf}) is even significantly larger than the term in Eq.\ (\ref{quantumBetaChemicalPotential}) because of $gn_0M/2k^2 \ll 1$.
Subsequently, the total effective chemical potential is negative for the droplets, as supported by the inset of Fig.~\ref{pressureDiagram}.

\subsection{Quartic exponent in the envelope\label{sec:quarticExponent}}

With the general formalism laid out in Sec.\ \ref{genericModel} we can now explore how the shape of the envelope affects the self-trapping mechanism of the cavity-induced quantum droplets.
To this end, we consider the same setup of Fig.~\ref{singleSetup} with a factorized envelope, but instead of Eq.\ (\ref{env}), we choose a quartic exponent according to
\begin{align}
f_{\xi}^{(4)}(\bm{r},\bm{r}') = e^{-(y^4+z^4)/\xi^4 - (y'^4+z'^4)/\xi^4} \, .
\label{singleCavityQuarticEnvelope}
\end{align}
For this, the inequality
\begin{align}
\bigg|\int_{-\frac{L}{2}}^{+\frac{L}{2}}\! \frac{dx}{L}\:\! e^{i\frac{2\pi m}{L}x} e^{-\frac{x^4}{\xi^4}} \:\!\bigg| < \frac{L^4}{4 \pi^2 m^2 \xi^4} \int_{-\frac{L}{2}}^{+\frac{L}{2}}\! \frac{dx}{L}\:\! e^{-\frac{x^4}{\xi^4}} \, ,
\label{quarticTransformationEstimate}
\end{align}
holds. We illustrate this inequality in Fig.\  \ref{graphicalVerification} (b) of Appendix \ref{inequalityVerification}. Thus, for $\bm{p},\bm{p}' \neq \bm{0}$ we obtain the estimate
\begin{align}
\hspace{-1mm}\frac{\big|\tilde{f}_{\xi}^{(4)} (\bm{p},\bm{p}')\big|}{\tilde{f}_{\xi}^{(4)} (\bm{0},\bm{0})} < \prod_{\nu=1,2}{}^{\!\!\!'} \frac{L^4}{4 \pi^2 m_\nu^2 \xi^4} \times \prod_{\nu=1,2}{}^{\!\!\!'} \frac{L^4}{4 \pi^2 m_\nu'^2 \xi^4} \, ,
\end{align}
where the primed products exclude the terms $m_\nu^{} = 0$ and $m_\nu' = 0$.
Taking into account that $(4\pi^2)^{-1} \approx 0.025$, we conclude that Eq.\ (\ref{approxKroneckerDelta}) is valid as long as $L < \xi$.
Consequently we only need to consider
\begin{align}
\tilde{f}_{\xi}^{(4)}(\bm{0},\bm{0}) = \left[ \frac{2\xi}{L} \Gamma\left( \frac{5}{4} \right) - \frac{\xi}{2L} \Gamma \left( \frac{1}{4} , \frac{L^4}{16\xi^4} \right) \right]^4 \, ,
\end{align}
where $\Gamma(s,x)$ is the upper incomplete gamma function and to find a roton given by Eq.\ (\ref{longRangeEigenmode}).
With this at hand we can investigate the effective energy $E_0$ in a similar manner as for the Gaussian envelope above.
\begin{figure}
\centering
\includegraphics[scale=.4]{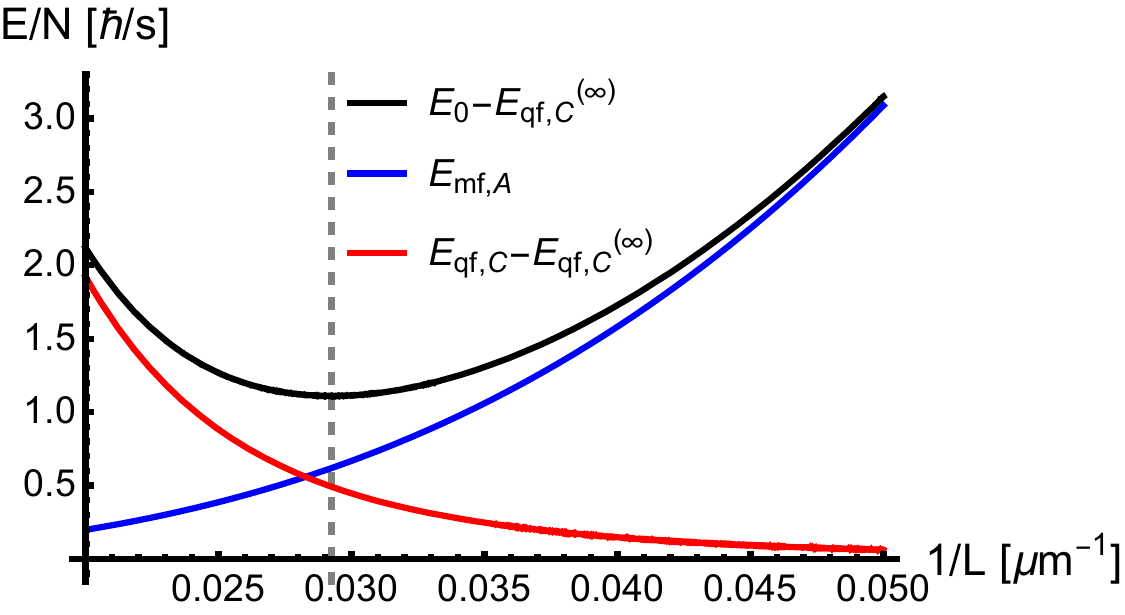}
\caption{Effective system energy $E_0$ per particle with its mean-field $E_{\rm mf,A}$ and quantum fluctuation cavity contribution $E_{\rm qf,C}$ for the quartic exponent envelope (\ref{singleCavityQuarticEnvelope}) in dependence of the inverse of the atomic system extension $L$.  A constant shift stemming from the infinite-range cavity $E_{\rm qf,C}^{(\infty)}$ has already been subtracted. An equilibrium droplet size $L_0$ is realized at the energy minimum marked by the gray dashed line. Remaining parameters are the same as Fig.~\ref{energyPerParticle}.  \label{quarticEnergyPotential}}
\end{figure}
Figure \ref{quarticEnergyPotential} shows the resulting 
energy minimum with respect to the system size.
The cavity correction shapes the effective energy potential as prescribed by its envelope.
Expanding the quartic envelope in Eq.\ (\ref{singleCavityQuarticEnvelope}) in the limit $L/\xi \to 0$ we find that the first nontrivial term is of the order $L^4/\xi^4$.  
Analogous to the case of the Gaussian envelope discussed above, this roton correction competes with the mean-field contact interaction.
However, we now get a weaker self-confinement of the system due to the different shape resulting from the quartic exponent. 
Consequently, the resulting droplet density depicted in Fig.~\ref{quarticEnergyPotential} is about $40 \, \%$ of the density realized with identical parameters for the Gaussian envelope in Fig.~\ref{energyPerParticle}.
The droplet class remains the same (D3), but now we have the exponent $\gamma = -7/3$.
This result indicates that a smaller exponent $\gamma$ results in a quantum droplet of larger density for otherwise identical parameters.
Adjusting the shape of the envelope is an intriguing tuning parameter in our setup.
By qualitatively changing the self-confinement it leads to a variation of the droplet size, the ground state energy and the parameter $\gamma$, as is further elaborated in the Appendix \ref{tuningExponent}.
\section{Translation-invariant interaction in a multi-mode cavity \label{translationInvariantMultiCavity}}
In the previous section we have analyzed the case of a single-mode cavity, where the envelope of the long-range interaction can be engineered by appropriately choosing the transversal modes of the cavity and the pump.
In the effective long-range interaction, Eq.\ (\ref{longRangePotential}), this results in an envelope $f_{\xi}(\bm{r},\bm{r}')$, which factorizes in its $\bm{r}$ and $\bm{r}'$ arguments as is exemplified, e.g., in Eq.\ (\ref{singleCavityQuarticEnvelope}).
A different situation was considered in Refs.\ \cite{kollar2015adjustable,kollar2017supermode,vaidya2018tunable,guo2019sign,guo2021optical}, which deals with an almost degenerate confocal cavity.
Thus, a multitude of cavity modes contributes to the effective long-range interaction, which can lead to a translation-invariant envelope in the plane orthogonal to the cavity axis.
In order to simplify the following calculation, we study the one-dimensional case in the pump direction $y$ of the effective interaction, which is analogous to the study of an external optical lattice with quantum Monte Carlo methods in Ref.~\cite{karpov2022light}.
By assuming that the atoms are placed in just one half-plane of the cavity at a distance much larger than the interaction range $\xi$, mirror image interactions are suppressed \cite{guo2021optical}.
Furthermore, an additional beam is used to cancel the non-translation invariant contributions in the interaction \cite{guo2021optical}, yielding an effective long-range interaction of the form
\begin{align}
V_{\rm C}(y,y') = \mathcal{I} \cos[k(y-y')] e^{-|y-y'|^2/\xi^2} \, .
\label{translationInvariantInteraction}
\end{align}
The total effective atom-only Hamiltonian then reads
\begin{eqnarray}
\hspace*{-2mm}&&\hat{H}_{\rm eff} = \int_{-\frac{L}{2}}^{+\frac{L}{2}} dy\, \hat{\psi}^{\dagger}(y) \left[ -\frac{\bm{\nabla}^2}{2M} + \frac{g}{2} \hat{\psi}^{\dagger}(y) \hat{\psi}(y)  \right] \hat{\psi}(y) \\
\hspace*{-2mm}&&{}+{} \frac{1}{2} \int_{-\frac{L}{2}}^{+\frac{L}{2}} dy \int_{-\frac{L}{2}}^{+\frac{L}{2}} dy'\, \hat{\psi}^{\dagger}(y) \hat{\psi}(y)V_{\rm C}(y,y') \hat{\psi}^{\dagger}(y') \hat{\psi}(y') \, .\nonumber 
\end{eqnarray}
Note that a similar effective system can be created in a ring cavity as was studied in the superradiant regime with mean-field methods in Ref.~\cite{masalaeva2023tuning}. In order to
check the condition \mbox{$\tilde{f}_{\bm{\xi}}({\bm p},{\bm p}') \ll \tilde{f}_{\bm{\xi}}(\bm{0},\bm{0})$} of the envelope, for \mbox{$\bm{m},\bm{m}' \in (\mathbb{Z} \setminus \{0\})^d$} for the envelope 
${f_{\xi}^{\rm (ti)}(y,y') = \exp(-|y-y'|^2/\xi^2)}$ we can apply the same methods as for a factorized envelope of the previous section.  However, here one  has to consider the special case $p=-p'$ in the same way as one has to analyze the situation $p=p'=0$ for a factorized envelope. 
The generic transformation
\begin{align}
\hspace*{-2mm}\tilde{f}_{\xi}^{\rm (ti)}(p,p') =\int_{-\frac{L}{2}}^{+\frac{L}{2}} \frac{dy}{L} \int_{-\frac{L}{2}}^{+\frac{L}{2}} \frac{dy'}{L} \:\! e^{-i(py+p'y')}e^{-\frac{|y-y'|^2}{\xi^2}} ,
\end{align}
for $p=-p'$ allows with the substitution $u=y'-y$ to obtain an estimate for the substituted integral by Eq.\ (\ref{singleTransformationEstimate}) and apply the approximation Eq.\ (\ref{approxKroneckerDelta}). This yields
\begin{align}
\tilde{f}_{\xi}^{\rm (ti)}(p,-p) \approx \delta_{p0} \tilde{f}_{\xi}^{\rm (ti)}(0,0) \, ,
\label{ti}
\end{align}
where the spatial average of the envelope occurring  in the roton mode reads
\begin{align}
\tilde{f}_{\xi}^{\rm (ti)}(0,0) = \frac{\xi^2}{L^2} \left( e^{-\frac{L^2}{\xi^2}} -1 \right) + \frac{\sqrt{\pi} \xi}{L} \,\text{erf} \left( \frac{L}{\xi} \right) \, .
\label{multiEnvelopeTransformation}
\end{align}
Analyzing Eq.\ (\ref{ti}) for  $p \neq -p'$ unveils that all terms in $\tilde{f}_{\xi}^{\rm (ti)}(p,p')$ are either suppressed polynomially like 
$1/(\pi m)^2$ or exponentially like $e^{-(\pi m L/\xi)^2}$, where $m \in \mathbb{Z} \setminus \lbrace 0 \rbrace$ is the integer determining the momentum $p = 2\pi m/L$. Thus we can apply once again Eq.\ (\ref{approxKroneckerDelta}) under the restriction $L/\xi < 1$. 

Based on these results for the transformation of the translation-invariant evenlope, we proceed with the Bogoliubov theory as outlined in Sec.\ \ref{genericModel}.
For the wavenumber $k\neq 0$ of the pump field the homogeneous mean-field energy turns out to be unaltered $E_{\rm mf,A} = g N^2/2L$.
Choosing the periodic function $v(y,y') = \cos[k(y-y')]$ in combination with the envelope $f_{\xi}^{\rm (ti)}(y,y') = \exp(-|y-y'|^2/\xi^2)$ leads to the translationally invariant long-range interaction of Eq.\ (\ref{translationInvariantInteraction}), which restricts the coupling between the modes to cases that are momentum conserving.
Thus, the Bogoliubov transformation leads to a fluctuation Hamilton operator of the generic form Eq.\ (\ref{h-fluc}) with an already  diagonal matrix $\underline{h}$. We then have
\begin{align}
\hspace*{-2mm}\hat{H}_2 ={}& \frac{1}{2} \sum_{p \notin \mathcal{K}_{\rm C}}{\!\!}^{'} \left( \hat{y}_{-p}\hat{y}_p + \omega_p^2 \hat{x}_{-p} \hat{x}_p - \frac{p^2}{2M} - gn \right) \nonumber\\
\hspace*{-2mm}&+ \frac{1}{2} \sum_{p = \pm k}{\!\!}^{'} \left( \hat{y}_{-p}\hat{y}_p +\Omega_p^2 \hat{x}_{-p} \hat{x}_p - \frac{p^2}{2M} - gn \right) \, ,\hspace*{-1mm}
\end{align}
with the roton dispersion following from Eq.\ (\ref{translationInvariantDispresion}) in the form 
\begin{align}
\Omega_p = \sqrt{\frac{p^2}{2M} \bigg[ \frac{p^2}{2M} + 2gn + \mathcal{I}N \tilde{f}_{\xi}^{\rm (ti)}(0,0) \bigg]} \, .
\end{align}
The roton formed here is part of the dispersion as visualized in Fig.~\ref{dispersion} in contrast to the previous cases of long-range interactions that are not translationally invariant.
Additionally, the roton correction involves here a roton at $p \in \{-k,k\}$ in this one-dimensional case. According to the translation-invariant variant of Eq.\ (\ref{longRangeCorrection}) this results in a prefactor of 2 in the one-dimensional cavity-induced quantum correction according to Eq.~\ref{longRangeZeroPointMotion}
\begin{align}
E_{\rm qf,C} = \Omega_k - \omega_k \, .
\label{multiCavityQuantumCorrection}
\end{align}
The effective energy
\begin{figure}
\centering
\includegraphics[scale=.4]{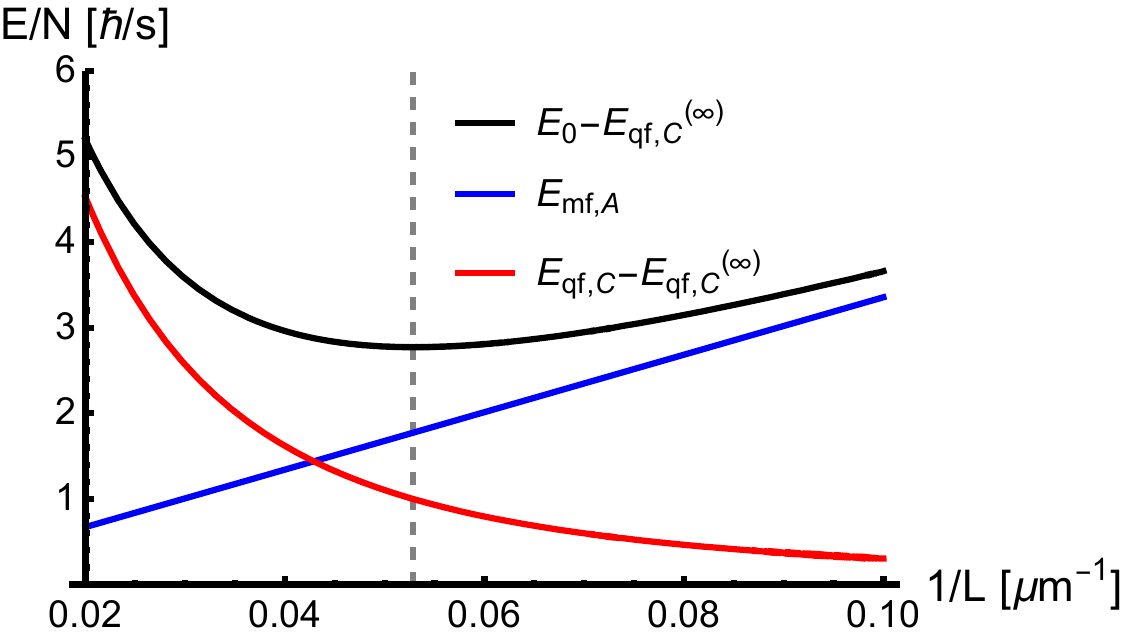}
\caption{Effective energy per particle and its constituents for the multi-mode cavity system realizing a translation-invariant long-range interaction versus the inverse system size $1/L$. The system-size independent energy shift of the infinite-range interaction term $E_{\rm qf,C}^{(\infty)}$ is subtracted. The droplet size $L_0$ at the minimum of the total system energy $E_0$ marked by the gray dashed line. Parameters are $k^2/2M = 2\pi \times 3560$ Hz, $\mathcal{I} = -21.25$ Hz, $N=10^3$, $\xi = 50 \, \mu$m, and $gn_0/(k^2/2M) = 1.6 \times 10^{-4}$. \label{multiEnergyPotential}}
\end{figure}
$E_0$ formed by the mean-field $E_{\rm mf,A} = g N^2/2L$ and the fluctuation correction of the translation invariant long-range interaction mediated by the multimode cavity from Eq.\ (\ref{multiCavityQuantumCorrection}) are depicted in Fig.~\ref{multiEnergyPotential}.
The repulsive mean-field energy $E_{\rm mf,A}$ scales with $1/L$.
For $\mathcal{I}<0$ the cavity-induced fluctuations cause the formation of two roton modes with negative quantum correction $E_{\rm qf,C}$.
As demonstrated in Fig.~\ref{multiEnergyPotential} it depends on the system size $L$ in such a way that it counters the mean-field energy.
Their competition forms a minimum of the effective energy $E_0$ at the equilibrium system size $L_0$.
Therefore, the system satisfies conditions (C1) and (C2).
Expanding the envelope transformation Eq.\ (\ref{multiEnvelopeTransformation}) around $\xi \to \infty$ according to
\begin{align}
\tilde{f}_{\xi}^{\rm (ti)}(0,0)= 1 - \frac{L^2}{6\xi^2} + \mathcal{O}\left( \frac{1}{\xi^4} \right) \, ,
\end{align}
reveals that the cavity-induced quantum fluctuation energy has again an infinite-range interaction term $E_{\rm qf,C}^{(\infty)}$ and a leading term of $L^2$ spatial dependence. Although the infinite-range interaction term does not play a role for conditions (C1) and (C2), it fixes the self-evaporation condition (C3). 
Thus, all three conditions for a quantum droplet are fulfilled by that translation-invariant system.
Applying the same approximations as for the single-mode cavity we find also for the multi-mode case a qualitative description of the effective energy in the form
\begin{align}
E_0(N,L) = E_{\rm qf,C}^{(\infty)} + \frac{gN^2}{2L} + \frac{D}{2} L^2 \, .
\end{align}
The prefactor of the quantum self-trapping of the system is given here by 
\begin{align}
D = \frac{-\mathcal{I}N}{6\xi^2\sqrt{1+(4gn+\mathcal{I}N)M/2k^2}} \, .
\end{align}
The qualitative dependence coincides with that of the two-dimensional factorized Gaussian envelope discussed in Sec.\ \ref{factorizedResults}. Quantitative differences result from the changed dimensionality and a twice as large prefactor in the quantum fluctuation correction caused by the translational invariance.
In comparison with the minimal model of Eq.\ (\ref{minimalModel}) we now consider a one-dimensional system $V=L$ with
the parameters  $\alpha = gN^2/2 >0$ and $\beta = D/2 >0$, which are basically unchanged compared to the factorized Gaussian envelope in a three-dimensional system, but the exponent is now different and amounts to $\gamma = - 3$.
Nonetheless, we have again a quantum droplet of class (D3) according to the general classification scheme in the introduction.
\section{Conclusions}\label{conclusions}
In this work, we study analytically a weakly interacting dilute Bose gas in a cavity.
Due to the cavity-induced long-range interaction, the Bose gas is effectively governed by two types of interactions with quite different ranges.
The cavity-induced interaction decays in space due to the mode envelope function and has a finite range that is much larger than the size of the atomic gas.
It creates a system size dependent coupling and effectively leads to the formation of a few distinct rotons whose depth depends on the system size.
Consequently, the energy correction resulting from the rotons varies with the size of the system and competes with the mean-field energy of the Bose gas, leading to the formation of a quantum droplet that replaces a simple BEC.
This mechanism for the emergence of quantum droplets differs significantly from the one already known for quantum droplets in dipolar Bose gases and Bose-Bose mixtures \cite{ferrier2016observation,cabrera2018quantum,semeghini2018self,skov2021observation}.
Namely, there a Feshbach resonance destabilizes the Bose gas in the mean field, while quantum fluctuations for contact or dipolar interaction then provide stabilization towards a liquid-like self-bound state.
Thus, we theoretically reveal an additional class of quantum droplets with unusual properties, which has not yet been considered before.
Since the destabilizing effect comes from only a few roton modes, the quantum fluctuation energy from the long-range interaction turns out to be not extensive.

These generic results are then specified for the well-established realization of the long-range interaction as an effective  coupling of atoms to a single mode of an optical cavity \cite{maschler2008ultracold,mivehvar2021cavity}. We found that the cavity-mediated interaction induces a roton mode whose properties can be modified by tuning cavity parameters as, e.g., the pump laser strength, the cavity detuning, or the waist of the pump beam. With this we can derive analytically the underlying effective energy whose extremalization yields the size of the self-bound quantum droplet. The corresponding predicted density turns out to be orders of magnitude more dilute in comparison with the already observed quantum droplets in dipolar Bose gases or Bose-Bose mixtures \cite{ferrier2016observation,cabrera2018quantum,semeghini2018self,skov2021observation}. To increase the quantum droplet density requires either a weaker contact interaction strength or a stronger cavity-mediated interaction strength. However, 
the latter is limited by the critical value for the self-organizing Dicke phase transition, where the homogeneous condensate breaks down.

From the point of view of a thermodynamic description the mean-field contribution of the contact interaction yields a positive pressure to the atomic system, while the quantum fluctuation correction of the cavity leads to a negative pressure.
From the resulting competition a mechanically stable quantum droplet emerges, whose positive bulk compressibility turns out to be smaller than that of a weakly interacting Bose gas without a cavity. Interestingly, the mechanical droplet criteria concerning pressure and compressibility are indifferent to a constant shift in the effective energy, which originates from an infinite-range interaction term. However, this term turns out to be indispensable to avoid a self-evaporation of the quantum droplet as it controls its chemical potential. With this also the thermodynamic investigation underlines that a Bose gas coupled to a single cavity mode leads to a novel droplet class. Furthermore, the fitting parameters of the effective energy potential, which determine the class of cavity-induced quantum droplets, can be modified by engineering both the extent and the spatial shape of the envelope that characterize the effective long-range interaction. 
This directly influences the properties of the quantum droplet such as, e.g., its size.
In particular, we found that a Gaussian envelope yields denser quantum droplets than a quartic envelope for otherwise same system parameters.

As a second special case investigated in this work is the realization of a translation-invariant long-range interaction, which can be engineered in multi-mode cavities. Our results show that the qualitative mechanism underlying the quantum droplet formation is the same as for the single-mode cavity, although the physical origin of the interaction envelope is quite different. Yet, quantitatively it turned out that the translation-invariant case has the tendency to lead to larger  quantum droplet densities as more roton modes contribute to that quantum fluctuation correction of the ground-state energy.
\section*{Acknowledgments}
We are grateful to Andreas Hemmerich and Hans Keßler for helpful discussions on the experimental parameters.
This work was supported by the Deutsche Forschungsgemeinschaft (DFG, German Research Foundation) via
the Collaborative Research Center SFB/TR185 (Project
No. 277625399) (A.P.) and via the Research Grant
274978739 (M.R. and M.T.). We also acknowledge the support from the DFG Cluster of Excellence CUI: ``Advanced Imaging of Matter'' -- EXC 2056 (Project ID 390715994).

\appendix
\onecolumngrid
\section{Numerical verification of the inequalities used \label{inequalityVerification}}

\begin{figure}[h]
    \centering    \includegraphics[width=.9\linewidth]{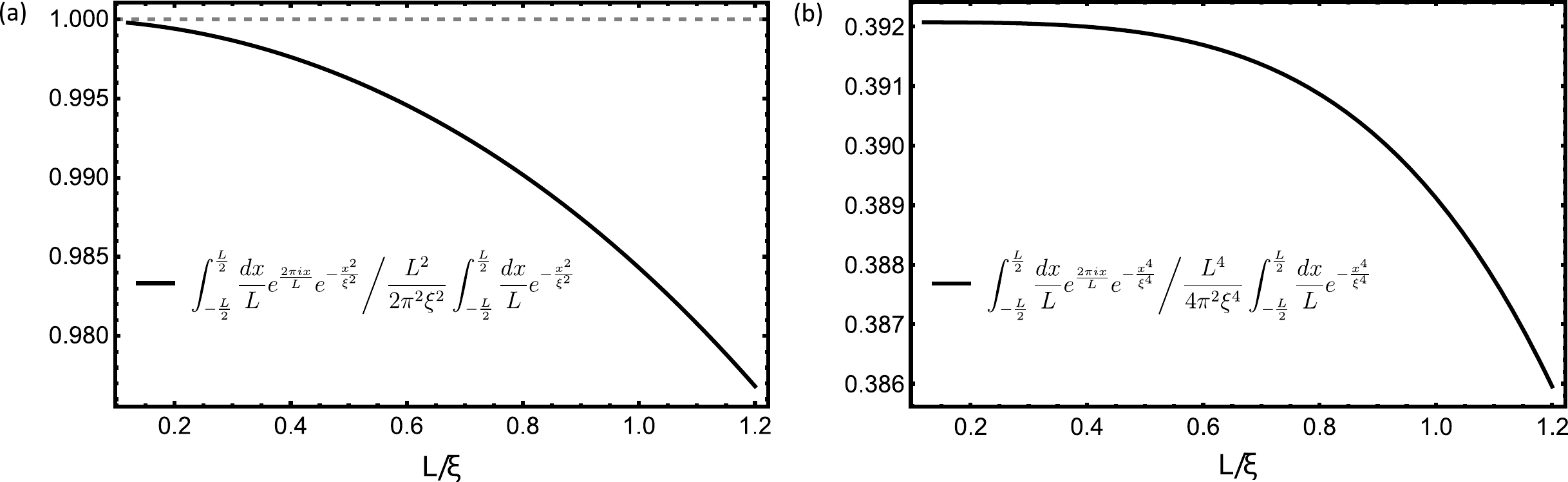}
    \caption{Graphical visualization of the validity of Eq.\ (\ref{singleTransformationEstimate}) in (a) and of Eq.\ (\ref{quarticTransformationEstimate}) in (b) in dependence of the system extension $L$ relative to the envelope width $\xi$. The horizontal line in (a) marks the value of $1$.}
    \label{graphicalVerification}
\end{figure}

Here, we illustrate in Fig.\ \ref{graphicalVerification} the validity of the inequalities Eqs.\ (\ref{singleTransformationEstimate}) and (\ref{quarticTransformationEstimate}) for $L < \xi$ by numerical evaluation.

\section{Tuning the exponent in the envelope\label{tuningExponent}}
\begin{figure}
\centering
\includegraphics[scale=.45]{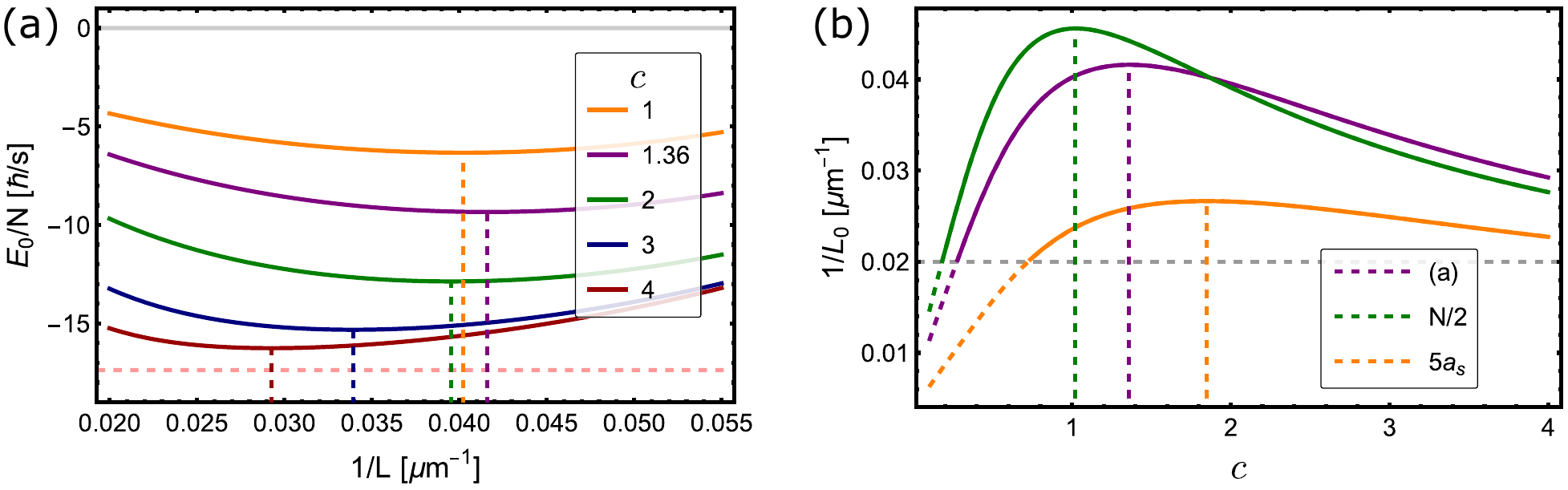}
\caption[Effective energy and droplet size under tunable envelope exponent]{In (a) the effective energy per particle $E_0/N$ is plotted against the inverse system size $1/L$ for five different choices of the envelope exponent $c$. The droplet size $L_0$ corresponding to the minimum of $E_0$ is indicated by a dashed line in the corresponding color.  The dashed horizontal line indicates the value of the infinite-range quantum correction $E_{\rm qf,C}^{(\infty)}/N$. The parameters are $N = 10^3$, $\xi = 50 \, \mu$m, $\mathcal{I} = -85$ Hz, $a_s = 100 \, a_0$, $M = 87$ u, $\omega_R = 2\pi \times 3560$ Hz.
In (b) we display the inverse droplet size $1/L_0$ against the exponent in the envelope $c$ for three parameter sets. The first, with the same parameters as in subplot (a), is labeled "{}(a)"{}. The set "{}$N/2$"{} uses identical parameters except that the number of atoms is halved to $N/2 = 500$. Finally, the parameter set "{}$5a_s$"{} uses the same parameters as subplot (a), but for five times the $s$-wave scattering length $5a_s = 500 \, a_0$.
The optimal exponent $c$ that leads to the densest droplet is indicated by a dashed vertical line. If $1/L_0 < 0.02 \, \mu{\rm m}^{-1}$, the condition  $L/\xi < 1$ is violated, which is necessary for the approximation leading to analytical solvability. To indicate this, a gray horizontal line separates the region of applicability above from the pathological region below. The curves are drawn as dashed lines in the region where the approximation breaks down to further indicate this. \label{variableEnvelope}}
\end{figure}
Here, we further explore the tuning and control of quantum droplet properties by introducing the exponent $c$ in the envelope with the generic ansatz
\begin{align}
f_{\xi}^{(c)}(\bm{r},\bm{r}') = \exp{\left(-\frac{|y|^{c}+|z|^{c}}{\xi^{c}}\right)} \exp{\left(-\frac{|y'|^{c}+|z'|^{c}}{\xi^{c}}\right)} \, .
\label{genericExponentEnvelope}
\end{align}
Assuming that the approximation Eq.\ (\ref{approxKroneckerDelta}) holds for an arbitrary $c > 0$ when the interaction is of global range $L<\xi$, we need only the spatial average
\begin{align}
\tilde{f}_{\xi}^{(c)}(\bm{0},\bm{0}) = \left\{ \frac{2 \xi}{c L}  \left[\Gamma \left(\frac{1}{c}\right)-\Gamma \left(\frac{1}{c},\left(\frac{L}{2\xi }\right)^{c}\right)\right] \right\}^4 = 1 - \frac{4}{1+c} \left( \frac{L}{2\xi} \right)^{c} + \mathcal{O}\left( \left( \frac{L}{2\xi} \right)^{2c} \right) \, .
\label{genericExponentSpatialAverage}
\end{align}
From Eq.\ (\ref{longRangeEigenmode}) the roton mode is determined by $\tilde{f}_{\xi}^{(c)}(\bm{0},\bm{0})$ with $\tilde{d} = 4$ and $\tilde{v} = 1/16$ and hence the quantum correction Eq.\ (\ref{longRangeZeroPointMotion}).
Analyzing the right-hand side of Eq.\ (\ref{genericExponentSpatialAverage}) we deduce that the first nontrivial term in the cavity-induced pressure is $P_{\rm qf,C} \propto L^{c-3}$.
We also find that the parameter $c$ is related to the droplet classification by $\gamma = -(1+c/3)$ for the three-dimensional system under investigation $V = L^3$. 
Hence the droplet is of type (D3) for any choice $c>0$.\\

We analyze the interplay between the Bose gas mean-field $E_{\rm mf,A}$ and the quantum correction $E_{\rm qf,C}$ as a function of the envelope exponent $c$ in Fig.~\ref{variableEnvelope}.
In panel (a) we find that the exponent parameter $c$ does affect the droplet size $L_0$. 
As suspected from Section \ref{sec:quarticExponent}, a larger exponent such as $c = 4$ results in a more dilute droplet than the Gaussian envelope $c = 2$.
In Fig.\ \ref{variableEnvelope} (a) we now find that decreasing the exponent below $c = 2$ leads, surprisingly, to a non-monotonic behaviour of the droplet density.
The maximum droplet density is obtained at the value $c = 1.36$ for the chosen parameter values.
This is further investigated in Fig.\ \ref{variableEnvelope} (b).
We observe that the optimal exponent $c$ to create the densest possible droplet is sensitive to the remaining set of parameters. 
In Fig.~\ref{variableEnvelope} (b), taking only half of the atoms $N/2$ leads to an optimal $c$ of about $1.02$.
Conversely, increasing the $s$-wave scattering by a factor of $5$ shifts the optimal exponent to $c \approx 1.85$.
In the limit $c \to 0$ the envelope Eq.\ (\ref{genericExponentEnvelope}) becomes spatially constant, so the interaction becomes infinite range.
Consequently, if $c \to 0$ a droplet can no longer be generated, so we obtain $1/L_0 \to 0$.\\

In Fig.\ \ref{variableEnvelope} (a) one can also see the relation between the effective energy per particle $E_0/N$ and the exponent $c$.
It is evident that a larger exponent leads to a more negative $E_0$. 
If we examine the right-hand side of Eq.\ (\ref{genericExponentSpatialAverage}), we see that the correction term $(L/2\xi)^{c}$ relative to the infinite-range order $1$ decreases in magnitude as the exponent $c$ increases.
Thus, a larger $c$ moves the effective energy $E_0$ closer to the value of $E_{\rm qf,C}^{(\infty)}/N$ ($\approx -17.37$ Hz for the chosen parameters), i.e., to more negative values.

\twocolumngrid

\end{document}